\documentclass[%
 reprint,
 amsmath,
 amssymb,
 prb,
 aps
]{revtex4-2}

\usepackage[english]{babel}

\usepackage{chemfig}
\usepackage{multirow}
\usepackage{graphicx}
\usepackage[
  colorlinks=true, 
  allcolors=blue
]{hyperref}
\usepackage{xcolor}
\usepackage{braket}
\usepackage{cleveref}
\usepackage[normalem]{ulem}

\begin{document}


\title{Dynamically Corrected Bethe--Salpeter Equation Solver for Self-consistent $GW$ Reference on the Matsubara Frequency Axis}

\author{Ming Wen}
\email{wenm@umich.edu}
\affiliation{Department of Chemistry, University of Michigan, Ann Arbor, MI 48109, USA}

\author{Gaurav Harsha}
\affiliation{Department of Chemistry, University of Michigan, Ann Arbor, MI 48109, USA}

\author{Dominika Zgid}
\affiliation{Department of Chemistry, University of Michigan, Ann Arbor, MI 48109, USA}
\affiliation{Department of Physics, University of Michigan, Ann Arbor, MI 48109, USA}
\affiliation{Faculty of Physics, University of Warsaw, Warsaw, Poland
\\}

\date{\today}

\begin{abstract}
We present a Bethe--Salpeter equation (BSE) solver based on a self-consistent $GW$ reference evaluated on the Matsubara frequency axis, referred to as BSE@sc$GW$.
The self-consistent $GW$ starting point provides a robust quasiparticle description and reduces sensitivity to the initial mean-field reference compared to one-shot $GW$-based approaches.
We further introduce a dynamical correction to the static Casida formulation via a plasmon-pole model.
This scheme incorporates simple dynamical screening effects while retaining the efficiency of an effective eigenvalue problem.
The resulting dynamically corrected BSE@sc$GW$ yields excitation energies in close agreement with high-level wavefunction-based benchmarks for both singlet and triplet excitations of small molecules. 
Overall, the accuracy of the dynamic BSE@sc$GW$ approach arises from the combination of a well-converged single-particle reference and the inclusion of frequency-dependent screening effects.
\end{abstract}

\maketitle


\section{Introduction}

Quantitatively reliable treatment of optical excitations,~\cite{blaseBetheSalpeterEquation2020} core excitations,~\cite{casanova-paezCoreExcitedStatesOpenShell2025} excitonic effects,~\cite{wangExcitonsSolidsPeriodic2020} and charge-transfer processes~\cite{mesterChargeTransferExcitationsDensity2022} requires advanced electronic-structure methods. 
Post-Hartree--Fock approaches such as coupled cluster (CC),~\cite{bartlettCoupledclusterTheoryQuantum2007a} configuration interaction (CI),~\cite{cremerConfigurationInteractionCoupled2013}
and equation-of-motion CC (EOM-CC),~\cite{krylovEquationofMotionCoupledClusterMethods2008b} provide systematically improvable and accurate excitation energies by incorporating electron correlation. However, their steep computational cost usually restricts them to finite molecular systems.

An alternative approach to excited state properties is the many-body perturbation theory (MBPT) expressed in the language of Green's functions (GFs).~\cite{fetterQuantumTheoryMany1971,aryasetiawanGWMethod1998, onidaElectronicExcitationsDensityfunctional2002a, reiningGWApproximationContent2018}
To treat neutral excitation processes using GF language, a two-particle bosonic GF is necessary, and such an approach is commonly formulated as the Bethe--Salpeter equation (BSE),~\cite{salpeterRelativisticEquationBoundState1951,strinatiEffectsDynamicalScreening1984,strinatiApplicationGreensFunctions1988} solved on top of a one-particle GF reference.
In principle, BSE depends on several mutually interdependent frequency (or time) variables, enabling the description of singlet, triplet, and double (HOMO$^2$-LUMO$^2$) excitations.~\cite{romanielloDoubleExcitationsFinite2009,sangalliDoubleExcitationsCorrelated2011}
While singlet and triplet states can be captured by a simplified particle-hole interaction kernel, the excitations with two-particle character can only be recovered with a full-frequency kernel treatment.~\cite{sangalliDoubleExcitationsCorrelated2011} 
For large molecular or periodic systems, a fully dynamical treatment is computationally demanding and often prohibitive. 
Consequently, a number of approximations are routinely introduced to simplify BSE, enabling practical implementations.~\cite{blaseBetheSalpeterEquation2020}

Customarily, the two-particle GF present in BSE is constructed from the one-particle GF. 
For this reason, BSE is commonly executed on top of a one-particle reference solution such as the $GW$ approximation.~\cite{baymConservationLawsCorrelation1961,baymSelfConsistentApproximationsManyBody1962,golzeGWCompendiumPractical2019a}
$GW$ approximation incorporates electron-electron correlation effects through self-energy.
The resulting Green’s function describes the propagation of an added or removed electron influenced by the many-electron environment.
When compared to the Hartree--Fock (HF) solution, $GW$ approximation leads to improved ionization/attachment energies~\cite{huserQuasiparticleGWCalculations2013,vansettenGW100BenchmarkingG0W02015b,govoniLargeScaleGW2015,maggioGWVertexCorrected2017a,wenComparingSelfConsistentGW2024} and refined band structures for solids.~\cite{godbySelfenergyOperatorsExchangecorrelation1988,hybertsenFirstPrinciplesTheoryQuasiparticles1985,hybertsenElectronCorrelationSemiconductors1986,garcia-gonzalezManyBody$mathitGW$Calculations2002,huserQuasiparticleGWCalculations2013,yehFullySelfconsistentFinitetemperature2022a}

Over the last few decades, the $GW$-based methods have seen rapid developments.~\cite{aryasetiawanGWMethod1998, reiningGWApproximationContent2018,golzeGWCompendiumPractical2019a}
The most commonly used variant is $G_0W_0$, which is a single-shot $GW$ method.~\cite{blaseFirstprinciples$mathitGW$Calculations2011,korzdorferStrategyFindingReliable2012,brunevalBenchmarkingStartingPoints2013b,vansettenGWMethodQuantumChemistry2013,vansettenGW100BenchmarkingG0W02015b} 
$G_0W_0$ uses the mean-field GF $G_0$ and the screened Coulomb interaction $W_0$ obtained from a single iteration to evaluate dynamical corrections for quasiparticle spectra.~\cite{hybertsenFirstPrinciplesTheoryQuasiparticles1985,hybertsenElectronCorrelationSemiconductors1986, godbySelfenergyOperatorsExchangecorrelation1988}
Beyond this heavily approximated scheme, the $GW$ method can be formulated at various levels of self-consistency, which improves conservation properties and reduces starting-point dependence, albeit at increased computational cost.~\cite{holmFullySelfconsistent$mathrmGW$1998b,vanschilfgaardeQuasiparticleSelfConsistent$GW$2006,shishkinSelfconsistent$GW$Calculations2007a,rostgaardFullySelfconsistentGW2010b,yehFullySelfconsistentFinitetemperature2022a}

While most early $GW$ implementations work on the real-frequency axis, $GW$ has also been formulated on the imaginary (Matsubara) frequency axis and the imaginary time axis, which are particularly well suited for finite-temperature and fully self-consistent implementations.~\cite{fetterQuantumTheoryMany1971,stanFullySelfconsistentGW2006,kutepovElectronicStructureNa2016b,kutepovSelfconsistentSolutionHedins2017,yehFullySelfconsistentFinitetemperature2022a}
By construction, the single-particle GFs can only describe charged excitations and falls short for neutral (or optical) excitations often relevant in two-body processes such as resonant photoemission spectroscopy.~\cite{onidaElectronicExcitationsDensityfunctional2002a,PhysRevB.100.085112}

The implementation of BSE is typically realized with three key simplifications:~\cite{CASIDA1996391,bechstedtCompensationDynamicalQuasiparticle1997,albrechtExcitonicEffectsOptical1998,onidaElectronicExcitationsDensityfunctional2002a,blaseBetheSalpeterEquation2018,choSimplifiedGWBSE2022,blaseBetheSalpeterEquation2020,yaoAllElectronBSEGW2022} 
(i) the electron-hole interaction kernel is treated as static, \textit{i.e.}, its frequency dependence is ignored, since dynamical effects in the screening of electron-hole processes and in single-particle processes are assumed to somewhat cancel each other out;
(ii) BSE inherits the widely used non-self-consistent, one-shot approximation in the underlying $GW$, resulting in the BSE@$G_0W_0$ scheme;
(iii) operationally, BSE is often cast into the Casida equation, which is also widely used in time-dependent Hartree--Fock (TD-HF) and time-dependent density functional theory (TD-DFT) methods.
Ultimately, BSE@$G_0W_0$ becomes a static eigenvalue problem after applying the three aforementioned simplifications.

BSE@$G_0W_0$ has emerged as a powerful and widely adopted method, especially in the condensed phase and material science communities.
It has been successful in accurately predicting excitation energies for atomic \textit{K}-edge excitations,~\cite{yaoAllElectronBSEGW2022} molecular neutral excitations,~\cite{blaseBetheSalpeterEquation2020,choSimplifiedGWBSE2022} and excitonic effects in strongly correlated solids.~\cite{delgrandeHowChooseEfficiently2025}
Nevertheless, some important limitations of the standard BSE@$G_0W_0$ protocol have been identified. 
First, its performance can degrade significantly for small molecules and localized excitations, where the approximations of a static kernel and one-shot quasiparticle corrections become more severe.~\cite{hiroseAllelectron$GW$+BetheSalpeterCalculations2015,blaseBetheSalpeterEquation2020} 
Second, because $G_0W_0$ is not self-consistent, the resulting quasiparticle energies, and hence the BSE excitation energies, are strongly dependent on the chosen starting point, \textit{i.e.} the mean field reference method.
This starting-point dependence can be exploited to optimize the mean-field reference, thereby can help achieve very accurate results.~\cite{gantOptimallyTunedStarting2022, wenComparingSelfConsistentGW2024}
Nonetheless, a manually chosen starting point is neither universally optimal nor appropriate for \textit{a priori} prediction.
In the absence of prior knowledge of the system or extensive benchmarking, the starting-point bias can also negatively affect results of the calculations.

To the best of our knowledge, there is little discussion about BSE based on fully self-consistent reference.~\cite{forsterQuasiparticleSelfConsistentGWBetheSalpeter2022}
Some efforts have been made to refine the basic approach of BSE@$G_0W_0$, such as using a partially self-consistent eigenvalue $GW$ (ev$GW$) reference.~\cite{knyshReferenceCC3Excitation2024}
The static approximation is another challenging avenue to pursue.
Strinati and coworkers reported the earliest exploration of dynamical corrections to BSE in periodic systems.~\cite{strinatiEffectsDynamicalScreening1984,strinatiApplicationGreensFunctions1988}
Building on this line of work, several groups have explored how to incorporate frequency-dependent effects into the BSE kernel in order to capture dynamical phenomena beyond a static screening approximation.~\cite{rohlfingElectronholeExcitationsOptical2000,maExcitedStatesBiological2009,zhangDynamicalSecondorderBetheSalpeter2013,authierDynamicalKernelsOptical2020,loosDynamicalCorrectionBethe2020,bintrimFullfrequencyDynamicalBethe2022}


In this study, we introduce an implementation of a dynamically corrected BSE built upon a fully self-consistent $GW$ reference on the imaginary frequency axis. 
To the best of our knowledge, a fully self-consistent $GW$ scheme on the Matsubara axis has not previously been integrated into the BSE framework.
We call this variant BSE@sc$GW$, where sc$GW$ stands for the starting point $GW$ being executed self-consistently.
The underlying sc$GW$ is performed using our recently developed, open-source finite-temperature implementation with Gaussian type orbital (GTO) basis sets.~\cite{iskakovGreenWeakCouplingImplementation2025}
Our BSE@sc$GW$ approach utilizes the frequency-dependent quantities obtained from the converged sc$GW$ calculations and recasts them into an effectively bosonized Hamiltonian using the Casida formalism. 
Consequently, beyond the static solution, we also compute dynamical corrections by employing a plasmon-pole fitting scheme.~\cite{hybertsenElectronCorrelationSemiconductors1986,larsonRolePlasmonpoleModel2013,golzeGWCompendiumPractical2019a}



\section{Theory}

\begin{figure*}
    \centering
    \includegraphics[width=0.9\linewidth]{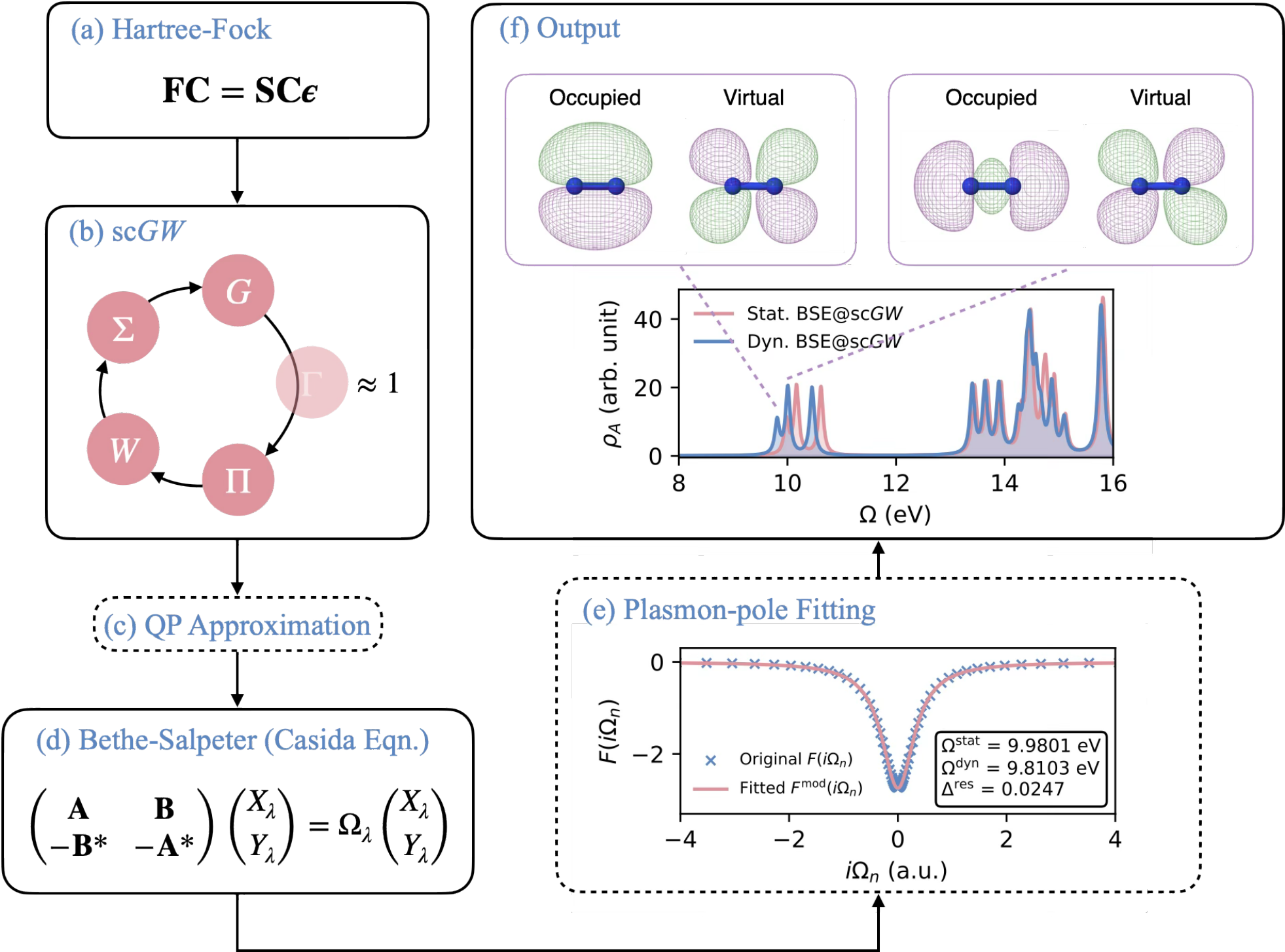}
    \caption{In counterclockwise order, the workflow of BSE@sc$GW$ proceeds as follows: (a) It begins with a mean-field calculation (HF in this study). (b) The density-fitted integrals and resulting matrices are then passed into the sc$GW$ cycle and iterated until self-consistency is reached. (c) The sc$GW$ quasiparticle energy levels are extracted using the QP approximation. (d) These QP energies, together with the corresponding output matrices, are subsequently used in the static BSE Casida equation. (e) The resulting static BSE solution is then dynamically corrected through plasmon-pole fitting. The shorthand ``a.u.'' stands for atomic unit. 
    (f) In the final step, the dynamically corrected excitation spectrum and the occupied-virtual MOs are generated and plotted. 
    For the static results, a default constant pole strength of 0.5 is used for each eigenvalue.
    The results shown in the two panels (e) and (f) on the right correspond to the N$_2$ molecule calculated with BSE@sc$GW$/aug-cc-pVTZ.}
    \label{fig:workflow}
\end{figure*}

In this section, we introduce  fundamental principles of our BSE@sc$GW$ approach, as depicted in FIG.~\ref{fig:workflow}.

\subsection{Self-consistent $GW$ and quasi-particle approximation}

Here, we briefly review the self-consistent implementation of $GW$ on the Matsubara frequency axis reported in previous work,~\cite{lanTestingSelfenergyEmbedding2017,iskakovInitioSelfenergyEmbedding2020,yehRelativisticSelfconsistent$GW$2022,yehFullySelfconsistentFinitetemperature2022a} which now is part of the \texttt{Green/WeakCoupling} 
package.~\cite{iskakovGreenWeakCouplingImplementation2025}
The $GW$ approach is derived from Hedin’s equations for many-body perturbation theory for interacting electron systems, which establish a closed set of relations between the self-energy, Green’s function, screened Coulomb interaction, irreducible polarization, and the vertex function.~\cite{hedinNewMethodCalculating1965}
The original equations of Hedin were formulated with the compact numerical labels, defined as 
\begin{equation}
\label{eq:num_label}
    1 \equiv (r_1, t_1),
\end{equation}
where $r_1$ is the real-space coordinate and $t_1$ stands for time.
The spin argument $\sigma_1$ is omitted.  
{Hedin’s equations represent diagrammatic relationships that are valid for both real- and imaginary-time variables.}

In this notation, the bare Coulomb operator $U(12)$ is defined in the density-density convention as
\begin{equation}
\label{eq:U_chem_notation}
\begin{split}
U(12) = \frac{1}{|r_1 - r_2|}.
\end{split}
\end{equation}
We note that in the four-point (orbital-basis) notation, the general two-body Coulomb matrix element reads
\begin{equation}
\begin{split}
    U_{ijkl} &\equiv ( ij | kl ) \\
    &= \iint dr_1 dr_2 \, \phi_i^*(r_1)\phi_j(r_1) \frac{1}{|r_1 -r_2|}\phi_k^*(r_2)\phi_l(r_2).
\end{split}
\end{equation}
where the ordering of orbital-basis indices follows the chemist notation $({ij|kl})$.
Using this notation,  the screened Coulomb interaction $W$ is calculated via a Dyson-like equation based on the bare Coulomb interaction $U$ as
\begin{equation}
\label{eq:real_W}
    W(12) = U(12) + W(13)\Pi(34)U(42), 
\end{equation}
where the irreducible polarizability $\Pi$ and vertex function $\Gamma$ are defined respectively as
\begin{equation}
    \Pi(12) = -iG(13)G(41)\Gamma(34;2),
\end{equation}
\begin{equation}
\label{eq: vertex}
    \Gamma(12;3) = \delta(13) \delta(23)+\frac{\delta \Sigma (12)}{\delta G(45)} G(46) G(75) \Gamma(67;3).
\end{equation}
In $GW$ without the vertex, higher-order corrections to the vertex function (the second term in Eqn.~\eqref{eq: vertex}) are ignored, resulting in $\Gamma(12;3) \approx \delta(13)\delta(23)$.~\cite{golzeGWCompendiumPractical2019a}
With this simplification, the computational cost for the self-energy $\Sigma$ and the polarization function $\Pi$ can be reduced. 
Their respective approximations read
\begin{subequations}
\begin{align}
\label{eq: GWA}
\begin{split}
\Sigma(12) &= iG(13)W(14)\Gamma(32;4) \\
&\approx iG(12)W(21^+),
\end{split}\\
\label{eq: Pi_12}
\begin{split}
\Pi(12) &= -iG(13)G(41)\Gamma(34;2) \\
&\approx -iG(12)G(21).
\end{split}
\end{align}
\end{subequations}
Correlated GF is then calculated with the non-interacting $G_0$ and self-energy as
\begin{equation}
\label{eq:real_dyson}
    G(12) = G_0(12) + G_0(13)\Sigma(34)G(42).
\end{equation}

In our finite-temperature $GW$ scheme, we reformulate all quantities appearing in Hedin’s equations on Matsubara-frequency axes (fermionic grid $i\omega_m$ and bosonic grid $i\Omega_n$), as well as on the imaginary-time axis $\tau$.
We also represent all quantities using the explicit atomic orbital labels ($i,j,k,l,$ \textit{etc.}) instead of the compact numeral labels, defined previously in Eqn.~\eqref{eq:num_label}.
The one-electron Matsubara GF on imaginary-time axis reads
\begin{equation}
\label{eq:G_tau_definition}
    G_{pq} (\tau) = -\frac{1}{\mathcal{Z}} \mathrm{Tr} \left[
        e^{-(\beta - \tau) (H - \mu N)} c_p e^{-\tau (H - \mu N)} c_q^\dagger
    \right],
\end{equation}
where $\mathcal Z $ is the grand-canonical partition function, ``$\mathrm{Tr}$'' denotes trace, $\beta$ is the inverse temperature, $\mu$ is the chemical potential, $c_p$ ($c_q^\dagger$) annihilates (creates) electrons in $p$-th ($q$-th) orbital, and $H$ and $N$ are the Hamiltonian and particle-number operators, respectively.
Note that when GF is defined intrinsically in imaginary time, no factor of $-i$ appears.~\cite{fetterQuantumTheoryMany1971, bruusManyBodyQuantumTheory2004} 
For further information on the Wick rotation in the complex time plane, refer to Appendix~\ref{sec: wick}.
The Matsubara GF $G(i\omega_n)$ and imaginary time GF $G(\tau)$ are related through Fourier transformation as
\begin{subequations}
    \begin{align}
        {G}(\tau) &= \frac{1}{\beta} \sum_m {G}(i\omega_m) e^{-i\omega_m \tau},
        \\
        {G}(i\omega_m) &= \int_0^\beta d\tau \, {G}(\tau) e^{i\omega_m \tau},
    \end{align}
\end{subequations}
where $i\omega_m$ denotes the fermionic Matsubara frequency. 

Transforming from space-time coordinates to orbital-based Matsubara representation,~\cite{yehFullySelfconsistentFinitetemperature2022a} the equations for the $GW$ approximation can be re-written as 
\begin{subequations}
\begin{align}
    \Pi_{abcd}(\tau) &= G_{da}(\tau)G_{bc}(-\tau),
    \label{eq:Pi_AO}
    \\
    \label{eq:coulomb_AO}
    \begin{split}
    W_{ijkl}(i\Omega_n) &= U_{ijkl} + \sum_{abcd} U_{ijab}\\ 
    &\quad \times \Pi_{abcd}(i\Omega_n) W_{cdkl}(i\Omega_n), 
    \end{split}
    \\
    \Sigma_{ij}(\tau) &= -\sum_{ab}G_{ab}(\tau)W_{iabj}(\tau^+).
    \label{eq:Sigma_AO}
\end{align}
\end{subequations}
The self-energy and GF are connected via the Dyson equation
\begin{equation}
    \boldsymbol{G}^{-1} (i\omega_n) = (i\omega_n + \mu) \mathbf{S} - \mathbf{H}_0 - \mathbf{\Sigma} (i\omega_n),
\label{eq:GF_AO}
\end{equation}
where $\mathbf{S}$ is the overlap matrix, $\mathbf{H}_0$ is the one-electron Hamiltonian describing the kinetic energy of electrons as well as their interaction with the nuclear charges.
Throughout this work, matrices are denoted in boldface, while individual matrix elements are written in regular type with subscripts.

In order to obtain the single-particle energy levels from GF, we employ the quasiparticle (QP) approximation~\cite{shishkinSelfconsistent$GW$Calculations2007a,harshaQuasiparticleFullySelfconsistent2024}.
Within the QP approach, we start from an effective one-body potential $V$ to provide an initial guess for quasiparticle eigenvalues $\epsilon_{p,0}$ and corresponding orbitals $\ket{\psi_p}$,
\begin{equation}
    \left(H_0 + V\right) \ket{\psi_p} = \epsilon_{p,0} \ket{\psi_p}.
\end{equation}
The $GW$-corrected quasiparticle energies are then obtained by replacing the contributions of $V$ with the self-energy $\Sigma$,
\begin{equation}
    \epsilon_p = \epsilon_{p,0} + \braket{\psi_p | \Sigma (\epsilon_p) - V | \psi_p},
    \label{eq:qp_energy}
\end{equation}
while still assuming $\ket{\psi_p}$ as approximate eigenvectors.
This is a non-linear equation as the self-energy is evaluated at $\epsilon_p$, which is the quasiparticle energy to be found.
In $G_0W_0$, the potential $V$ and initial $\epsilon_{p,0}$ come from the initial DFT or HF solution. 
In our fully self-consistent $GW$, we instead use the converged static self-energy to get the initial input $\Sigma$ for this purpose.

\begin{figure}
    \centering
    \includegraphics[width=0.95\linewidth]{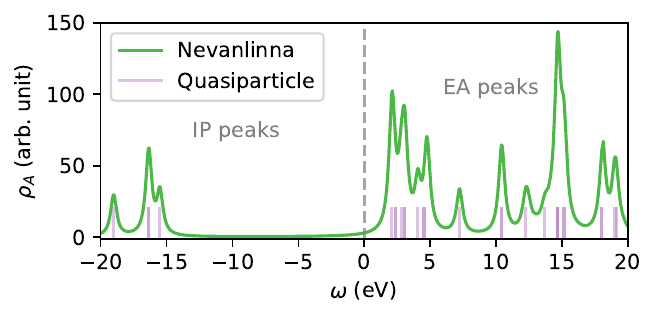}
    \caption{Charged excitations of N$_2$ calculated at the sc$GW$/aug-cc-pVTZ level of theory. 
    The negative and positive half-axes contain the ionization potential (IP) peaks and  electronic affinity (EA) peaks respectively.
    The QP energy levels serve as the initial inputs for the subsequent BSE calculation shown in FIG.~\ref{fig:workflow} (e) and (f).
    }
    \label{fig:QP vs NAC}
\end{figure}

We note that, in contrast to quasiparticle self-consistent $GW$ (qs$GW$) approaches,~\cite{vanschilfgaardeQuasiparticleSelfConsistent$GW$2006,kutepovElectronicStructureNa2016b} this QP approximation merely serves as a numerically robust alternative to analytic continuation for obtaining quasiparticle energies that enter BSE.
This is reaffirmed in FIG.~\ref{fig:QP vs NAC} where we compare Nevanlinna analytic continuation~\cite{feiNevanlinnaAnalyticalContinuation2021a} results and QP energies obtained from Eqn.~\eqref{eq:qp_energy} for $N_2$ molecule.
It is evident that both these approaches result in almost identical energies for charged excitations, particularly near the fermi level, justifying the use of QP approach for BSE calculations based on sc$GW$.

\subsection{Bethe--Salpeter equation}

The one-particle GF characterizes single-particle processes associated with quantities such as the ionization potential ($E_{\mathrm{IP}}$), electron affinity ($E_{\mathrm{EA}}$), and fundamental gap ($\Delta E_g = E_{\mathrm{EA}} - E_{\mathrm{IP}}$). 
A variety of $GW$-based approaches, including $G_0W_0$,~\cite{blaseFirstprinciples$mathitGW$Calculations2011,korzdorferStrategyFindingReliable2012,brunevalBenchmarkingStartingPoints2013b,vansettenGWMethodQuantumChemistry2013,vansettenGW100BenchmarkingG0W02015b} vertex corrected $GW$ ($GW\Gamma$),~\cite{ren$GW$ApproximationSecondorder2015,knightAccurateIonizationPotentials2016b,lewisVertexCorrectionsPolarizability2019b,vlcekStochasticVertexCorrections2019a,maggioGWVertexCorrected2017a,wangAssessingG0W0G01Approach2021,mejuto-zaeraAreMultiquasiparticleInteractions2021a} and fully or partial self-consistent $GW$ (sc$GW$)~\cite{holmFullySelfconsistent$mathrmGW$1998b,rostgaardFullySelfconsistentGW2010b,brunevalIonizationEnergyAtoms2012,knightAccurateIonizationPotentials2016b,carusoBenchmarkGWApproaches2016,rangelEvaluatingGWApproximation2016a,kaplanQuasiParticleSelfConsistentGW2016a,wenComparingSelfConsistentGW2024} have been successfully used to predict these properties.
In contrast, most types of molecular spectroscopy, including electron energy loss spectroscopy (EELS), are sensitive to optical excitations.~\cite{onidaElectronicExcitationsDensityfunctional2002a}
Such optical excitation processes are referred to as ``neutral excitation'' since they arise due to a redistribution of electrons when compared with the ground state (or a parent state). 

\begin{figure}
    \centering
    \includegraphics[width=\linewidth]{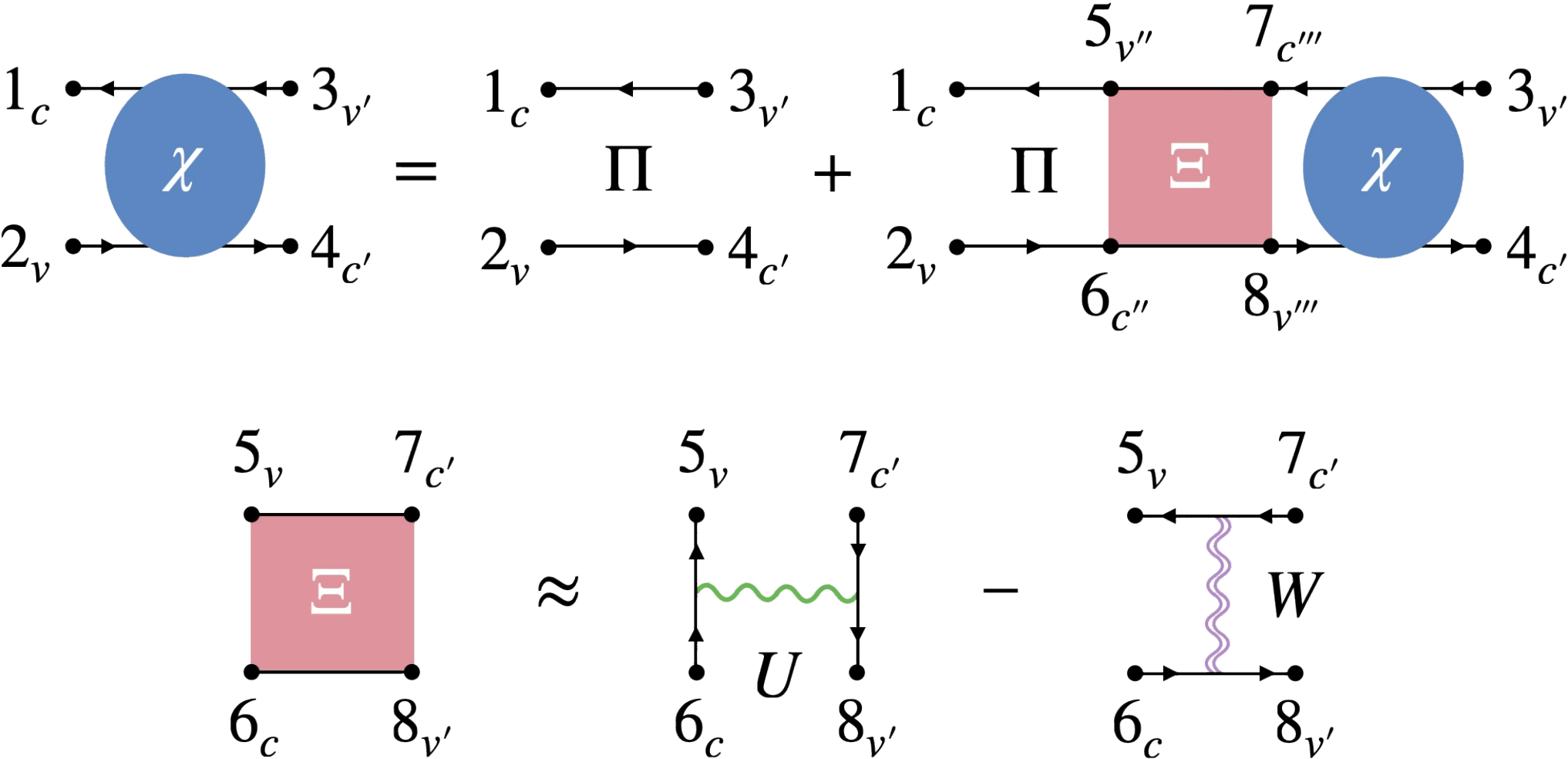}
    \caption{Feynman diagram of the particle-hole Bethe--Salpeter equation (see Eqn.~\eqref{eq:BSE}). 
    Subscripts signify valence and conduction orbitals.
    BSE kernel $\Xi$ is approximated as the sum of the electron-hole exchange term $U$ and the electron-hole attraction term $-W$ in Eqn.~\eqref{eq: kernel_omega}.
    }
    \label{fig:BSE_Feynman}
\end{figure}

To describe such electron redistribution processes, a two-particle correlation function needs to be introduced.
Formally, the two-particle correlation function is a four-point susceptibility, defined in the BSE formalism as
\begin{equation}
    \chi(12;34) = \frac{\delta G(12)}{\delta \phi(34)}, 
\end{equation}
where $\phi$ is an external non-local perturbation.~\cite{strinatiApplicationGreensFunctions1988}
The functional derivative $\frac{\delta G}{\delta \phi}$ captures the linear response, where correlations between two space-time coordinates at (1,2) are generated by the infinitesimal perturbation at (3,4). 
In the non-interacting limit, $\chi$ reduces to $\Pi$ as in Eqn.~\eqref{eq: Pi_12}.
Pristine BSE is given as a Dyson-like equation involving the two-particle correlation functions as
\begin{equation}
\label{eq:BSE}
    \chi(12;34) = \Pi(12;34) + \Pi(12;56) \Xi(56;78) \chi(78;34). 
\end{equation}
The Feynman diagrams for BSE are shown in FIG.~\ref{fig:BSE_Feynman}.
{Analogous to Hedin’s equations, BSE represents the diagrammatic relationship among two-particle correlation functions, formulated for a general time argument and not limited to either real or imaginary time. }
BSE relates the non-interacting particle-hole polarization $\Pi$ to the fully interacting particle-hole polarization $\chi$, 
representing an infinite series of ladder diagrams mediated by the kernel $\Xi$.
In the remainder of this section, we outline the theoretical foundations that support the BSE implementation used in this study.
The complete workflow is depicted in FIG.~\ref{fig:workflow}.

We restrict ourselves to using only imaginary times as arguments ($1 \equiv r_1, \tau_1$) since we are working with Matsubara Green’s functions.
The four-point response function $\chi(12;34)$ and the BSE kernel $\Xi(56;78)$ are projected onto the occupied-virtual molecular orbital basis by integrating out the spatial degrees of freedom as in Eqn.~\eqref{eq: real_to_orb}.
Indices $c$ and $v$ label conduction (virtual) and valence (occupied) states, respectively.
\begin{widetext}
\begin{subequations}
\label{eq: real_to_orb}
\begin{align}
    \chi_{cvc'v'}(\tau_1,\tau_2,\tau_4,\tau_3) &=
    \iiiint  dr_1\,dr_2\,dr_3\,dr_4\; \phi^*_c(r_1) \phi_{v}(r_2)\chi(12;34)\phi^*_{c'}(r_4)\phi_{v'}(r_3), \\
    \Xi_{vcv'c'}(\tau_5,\tau_6,\tau_8,\tau_7) &= \iiiint  dr_5\,dr_6\,dr_7\,dr_8\;\phi^*_{v}(r_5)\phi_{c}(r_6)\Xi(56;78)\phi^*_{v'}(r_8)\phi_{c'}(r_7).
\end{align}
\end{subequations}
\begin{equation}
\begin{split}
\label{eq:BSE_AO}
\chi_{cvc'v'}(\tau_1,\tau_2,\tau_4,\tau_3)
&= \Pi_{cvc'v'}(\tau_1,\tau_2,\tau_4,\tau_3) + \sum_{v''v'''c''c'''} \bigg[
\iiiint d\tau_5\, d\tau_6 \, d\tau_7 \, d\tau_8 \; \Pi_{cvc''v''}(\tau_1,\tau_2,\tau_6,\tau_5)  \\
&\quad  \times
\Xi_{v''c''v'''c'''}(\tau_5,\tau_6,\tau_8,\tau_7)\,
\chi_{c'''v'''c'v'}(\tau_7,\tau_8,\tau_4,\tau_3)\bigg].
\end{split}
\end{equation}
\end{widetext}
The kernel $\Xi$ has the following definition as a functional derivative
\begin{equation}
\Xi_{vcv'c'}(\tau_5,\tau_6,\tau_8,\tau_7) = 
\frac{\delta [V^{H}_{vc}(\tau_5)\delta(\tau_5-\tau_6) + \Sigma_{vc}(\tau_5,\tau_6)]}{\delta G_{v'c'}(\tau_8,\tau_7)}.
\end{equation}
The first term is the functional derivative of the Hartree potential $V^H$ w.r.t. GF, which reads
\begin{equation}
\label{eq:kernel_1st_term}
\begin{split}
&\frac{\delta [V^{H}_{vc}(\tau_5)\delta(\tau_5-\tau_6)]}{\delta G_{v'c'}(\tau_8,\tau_7)} = \\
&\quad U_{vcv'c'}\bigg[\delta(\tau_5-\tau_6)\,\delta(\tau_5-\tau_8)\,\delta(\tau_5-\tau_7)\bigg].
\end{split}
\end{equation}
The second term is more complicated. 
Firstly, we approximate the self-energy as the $GW$ self-energy $\Sigma^{GW}$, which gives
\begin{equation}
\label{eq:sigma_to_G}
\begin{split}
&\frac{\delta\Sigma_{vc}(\tau_5,\tau_6)}{\delta G_{v'c'}(\tau_8,\tau_7)} 
\approx \frac{\delta\Sigma^{GW}_{vc}(\tau_5,\tau_6)}{\delta G_{v'c'}(\tau_8,\tau_7)} \\
&= \frac{\delta \left[
-\sum_{rs} G_{rs}(\tau_5,\tau_6)\,W_{vrsc}(\tau_5-\tau_6)
\right]}{\delta G_{v'c'}(\tau_8,\tau_7)}\\
&= -\sum_{rs} 
\frac{\delta G_{rs}(\tau_5,\tau_6)}{\delta G_{v'c'}(\tau_8,\tau_7)}
W_{vrsc}(\tau_5-\tau_6) \\
&\quad  -\sum_{rs} G_{rs}(\tau_5,\tau_6)
\frac{\delta W_{vrsc}(\tau_5-\tau_6)}{\delta G_{v'c'}(\tau_8,\tau_7)}, 
\end{split}
\end{equation}
where the indices ($r,s$) are dummy orbital indices used for contraction.
It is a common practice in BSE to eliminate the functional derivative $\frac{\delta W}{\delta G}$ by simply approximating it as zero. 
\footnote{Note: However, there is no clear physical justification for dropping the term containing $\frac{\delta W}{\delta G}$. 
Rohfling and Louie~\cite{rohlfingElectronholeExcitationsOptical2000} have emphasized that $\frac{\delta W}{\delta G}$ is extremely challenging to evaluate.}
With $\frac{\delta W}{\delta G} = 0$, Eqn.~\eqref{eq:sigma_to_G} becomes
\begin{equation}
\label{eq:kernel_2nd_term}
\begin{split}
&\frac{\delta\Sigma_{vc}(\tau_5,\tau_6)}{\delta G_{v'c'}(\tau_8,\tau_7)} 
\approx -\sum_{rs} 
\frac{\delta G_{rs}(\tau_5,\tau_6)}{\delta G_{v'c'}(\tau_8,\tau_7)}
W_{vrsc}(\tau_5-\tau_6) \\
&= -\sum_{rs} 
\delta_{rv'}\,\delta_{sc'}\,W_{vrsc}(\tau_5-\tau_6)\bigg[\delta(\tau_5-\tau_8)\,\delta(\tau_6-\tau_7)\bigg] \\
&= -W_{vv'c'c}(\tau_5-\tau_6)\bigg[\delta(\tau_5-\tau_8)\,\delta(\tau_6-\tau_7)\bigg].
\end{split}
\end{equation}
Combining Eqns.~\eqref{eq:kernel_1st_term} and \eqref{eq:kernel_2nd_term}, the approximated BSE kernel is expressed as
\begin{equation}
\label{eq:kernel_explicit_tau}
\begin{split}
\Xi_{vcv'c'}&(\tau_5,\tau_6,\tau_8,\tau_7) 
= \\
& U_{vcv'c'}\bigg[\delta(\tau_5-\tau_6)\,\delta(\tau_5-\tau_8)\,\delta(\tau_5-\tau_7)\bigg]\\ 
- &W_{vv'c'c}(\tau_5-\tau_6)\bigg[\delta(\tau_5-\tau_8)\,\delta(\tau_6-\tau_7)\bigg].
\end{split}
\end{equation}
It is useful to define the time relations among ($\tau_5$,$\tau_6$,$\tau_7$,$\tau_8$) in order to eliminate multiple time Kronecker delta functions appearing in Eqn.~\eqref{eq:kernel_explicit_tau}. 
We therefore invoke the following approximation: 
the particle-hole transferring excitation is relatively long-lived; while the single-particle processes (creation and annihilation) are instantaneous. 
Consequently, the single-particle GFs of an electron and a hole entering the particle-hole bubble are replaced by their quasiparticle approximated spectral forms.~\cite{rohlfingElectronholeExcitationsOptical2000} 
Under this approximation, the fermionic frequency sums involve only simple pole structures and can be performed analytically, independently of the bosonic frequency.
Based on this rationale, we introduce the relative bosonic time $T$ in
\begin{equation}
\begin{cases}
\tau_5 = \tau_8, \\
\tau_6 = \tau_7, \\
T = \tau_5 - \tau_6.
\end{cases}
\end{equation}
In the electron-hole excitation process, $\tau_5 = \tau_8$ indicates that an electron is created and annihilated instantaneously, and $\tau_6 = \tau_7$ indicates that a hole is created and annihilated instantaneously.
The difference $T$ is referred to as the relative time of the electron-hole pair, or exciton. 

Energy conservation dictates that the two-particle correlation function depends only on $T$ rather than on $\tau_5$ and $\tau_6$ independently. 
$\chi(T)$ is also symmetric under bosonic statistics, this dependence on the time difference allows one to define a Fourier transform leading to the bosonic frequency $i\Omega_n$.
The two fermionic Matsubara frequencies $i\omega_m$ and $i\omega_{m'}$ describe the two propagations within each particle-hole pair, while the transferred bosonic frequency is the difference as
\begin{equation}
i\Omega_n = i\omega_m - i\omega_{m'}.
\end{equation}

Using this relation, Eqn.~\eqref{eq:kernel_explicit_tau} now only depends on a single bosonic time argument $T$ as
\begin{equation}
\label{eq:kernel_T}
\Xi_{vcv'c'}(T) = U_{vcv'c'}\delta(T) - W_{vv'c'c}(T).
\end{equation}
The Kronecker delta function $\delta(T)$ naturally arises in the first term. 
It corresponds to $U$ being an instantaneous quantity in time. 
The Fourier transformation of Eqn.~\eqref{eq:kernel_T} gives the frequency-dependent form of $\Xi$ as
\begin{equation}
\label{eq: kernel_omega}
\Xi_{vcv'c'}(i\Omega_n) = U_{vcv'c'} - W_{vv'c'c}(i\Omega_n).
\end{equation}
Note that the kernel's frequency dependence enters only through the screened Coulomb interaction term $W$. 
The final form of the BSE kernel corresponds to the diagram given in FIG.~\ref{fig:BSE_Feynman}. 
It can be viewed diagrammatically as the sum of an exchange term and attraction term connecting two electron-hole propagation lines.  

\subsection{Density fitting}
In our self-consistent $GW$ implementation, we use a density-fitted, or resolution of identity (DF-RI) two-electron integral~\cite{dunlapRobustVariationalFitting2000,wernerFastLinearScaling2003,renResolutionofidentityApproachHartree2012,yeFastPeriodicGaussian2021} to reduce computation and memory cost. 
$U_{ijkl}$ represents the 4-dimensional bare two-electron Coulomb interaction, which can be decomposed into 3-dimensional tensors $V_{ij,Q}$ with the help of an auxiliary basis as
\begin{equation}
    U_{ijkl} \equiv  \sum_Q V_{ij,Q}V_{kl,Q},
\end{equation}
where the auxiliary basis is indexed by $Q$.
Using this decomposition, the screened interaction $W_{ikjl}(i\Omega_n)$ can be represented as
\begin{equation}
\begin{split}
W_{iklj}(i\Omega_n) &\equiv \sum_Q V_{ik,Q}V_{lj,Q} \\
&\quad + \sum_{QQ'} V_{ik,Q}\tilde P_{QQ'}(i\Omega_n)V_{lj,Q'},
\end{split}
\end{equation}
where the two-point polarization $\tilde P_{QQ'}(i\Omega_n)$ is expressed using the auxiliary basis.
Finally, inserting such an expression for $W$ into the kernel $\Xi_{ijkl}(i\Omega_n)$ equation yields:
\begin{equation}
\begin{split}
    \Xi_{ijkl}(i\Omega_n) = &\sum_Q [V_{ij,Q}V_{kl,Q} - V_{ik,Q}V_{lj,Q}] \\
    &- \sum_{QQ'} V_{ik,Q}\tilde P_{QQ'}(i\Omega_n)V_{lj,Q'}.
\end{split}
\end{equation}
This final expression provides a compact orbital representation of the Bethe--Salpeter kernel, where the frequency dependence is embedded in the auxiliary-basis polarization $\tilde P_{QQ'}(i\Omega_n)$. 
This quantity requires significantly less storage than the complete four-point objects.

\subsection{Casida equation formalism}
A practical way to solve BSE is to recast the particle-hole polarization in Eqn.~\eqref{eq:BSE} into a generalized Hamiltonian eigenvalue problem, which resembles the Casida equation found in the TD-HF and TD-DFT methods.~\cite{onidaElectronicExcitationsDensityfunctional2002a} 
The Casida equation is formulated as
\begin{equation}
\label{eq: casida_OG}
\begin{pmatrix}
\textbf{A} & \textbf{B} \\ 
-\textbf{B}^* & -\textbf{A}^* 
\end{pmatrix}\begin{pmatrix}
X_\lambda \\ 
Y_\lambda
\end{pmatrix} = \Omega_\lambda\begin{pmatrix}
X_\lambda \\ 
Y_\lambda
\end{pmatrix}.
\end{equation}
The blocks $\textbf{A}$ and $\textbf{B}$, defined in the occupied-virtual MO 
space with the compact index $(vc)$, are
\begin{subequations}
\label{eq:casida-AB-defs}
\begin{align}
\Delta \epsilon_{vcv'c'} &= (\epsilon_c - \epsilon_v)\delta_{vv'}\delta_{cc'},  \\
\label{eq: block A}
A_{(vc)(v'c')} &= \Delta \epsilon_{vcv'c'} +\kappa \, U_{vcv'c'} - W_{vv'c'c},  \\
\label{eq: block B}
B_{(vc)(v'c')} &= \kappa \, U_{vcc'v'} - W_{vc'v'c}. 
\end{align}
\end{subequations}
where $\Delta \epsilon$ uses sc$GW$ quasiparticle energies, $U$ is 
reconstructed via DF-RI, and $W$ is taken from the converged sc$GW$ solution. 
Each block has dimension $(n_v n_c \times n_v n_c)$, giving an effective 
Hamiltonian of size $(2n_vn_c \times 2n_vn_c)$.

Starting from a closed-shell spin-restricted case, the BSE kernel $\Xi$ can be decoupled to account for singlet and triplet excitations explicitly
through a parameter $\kappa$ in Eqn.~\eqref{eq:casida-AB-defs},~\cite{rohlfingElectronHoleExcitationsSemiconductors1998,rohlfingElectronholeExcitationsOptical2000,onidaElectronicExcitationsDensityfunctional2002a} defined as
\begin{equation}
\label{eq: kappa}
\kappa = \begin{cases}
    0 \mathrm{\ for\ triplets}, \\
    2 \mathrm{\ for\ singlets}.
\end{cases}
\end{equation}
This decoupling relies on neglecting spin-orbit coupling, which we adopt for the entirety of this study.

Note that a non-empty coupling block $\textbf{B}$ results in a non-Hermitian Hamiltonian. 
A common approach to restore Hermiticity is the Tamm-Dancoff approximation (TDA) which sets $\mathbf{B} = \textbf{0}$.~\cite{hirataTimedependentDensityFunctional1999}
We retain the full kernel with $\textbf{B}$ throughout this work.

Crucially, $W$ in Eqn.~\eqref{eq:casida-AB-defs} is frequency-dependent. 
The standard treatment invokes the static approximation, 
$W(\Omega) \approx W(\Omega = 0)$, replacing the screened interaction by its zero-frequency limit. 
The consequences of going beyond this approximation are addressed in the following section.

\subsection{Dynamical effective Hamiltonian and plasmon-pole fitting}
The static approximation discards the frequency dependence of $W$ and potentially misses dynamical correlation effects in the excitation spectrum. 
Several methods have been proposed to remedy this on the real frequency axis.~\cite{romanielloDoubleExcitationsFinite2009,blaseBetheSalpeterEquation2020,loosDynamicalCorrectionBethe2020,loosStaticDynamicBethe2022} 

Here, we follow the same philosophy on the imaginary Matsubara axis, formulating BSE as a non-linear eigenvalue problem with a frequency-dependent effective Hamiltonian:
\begin{equation}
\textbf{H}^{\mathrm{eff}}(i\Omega_n) =\begin{pmatrix}
\textbf{A}(i\Omega_n) & \textbf{B}(i\Omega_n) \\ 
-\textbf{B}^*(i\Omega_n) & -\textbf{A}^*(i\Omega_n) 
\end{pmatrix}. 
\end{equation}
We essentially treat Eqn.~\eqref{eq: casida_OG} as a non-linear eigenvalue problem. 
First the eigenvalue equation is solved with the static approximation $\textbf{H}^{\mathrm{stat}} \equiv \textbf{H}^{\mathrm{eff}}(i\Omega_n = 0)$ as
\begin{equation}
\label{eq: static eig eq}
\textbf{H}^{\mathrm{stat}} \textbf{V}=\textbf{V} \boldsymbol{\Lambda}^{\mathrm{stat}},
\end{equation}
where $\boldsymbol{\Lambda}^{\mathrm{stat}}$ reproduces the standard static BSE solution.
To avoid re-diagonalizing at every frequency point, we adopt an adiabatic approximation: the eigenvector matrix 
$\textbf{V}$ from the static problem is assumed to diagonalize 
$\textbf{H}^{\mathrm{eff}}(i\Omega_n)$ at all frequencies,
\begin{equation}
{\bf \Lambda}(i\Omega_n) \approx \textbf{V}^{-1} \textbf{H}^{\mathrm{eff}} (i\Omega_n)\textbf{V}.
\end{equation}
This reduces storage of dense matrices to a set of $n_{\mathrm{freq}}$ diagonal matrices. 
However, discarding the off-diagonal elements of $\textbf{V}^{-1}\textbf{H}^{\mathrm{dyn}}(i\Omega_n)\textbf{V}$ introduces a diagonalization error. 

Similarly to the definition of single-particle GF,
an auxiliary response function can be constructed from ${\bf \Lambda}(i\Omega_n)$ as:
\begin{equation}
\label{eq: response to Heff}
{\bf F}(i\Omega_n) \equiv \int d\omega\, \frac{\boldsymbol{\rho}_A(\Omega)}{i\Omega_n -\Omega} = \frac{1}{i\Omega_n \cdot {\bf I} - {\bf \Lambda}(i\Omega_n)}.
\end{equation}
whose spectral poles in $\rho_A(\omega)$ yield the particle-hole excitation energies.

Although the adiabatic approximation removes dynamical off-diagonal coupling, the analytic continuation of $\mathbf{F}(i\Omega_n)$ from the imaginary to the real frequency axis remains non-trivial.
We therefore adopt a physically motivated plasmon-pole approximation, as illustrated in FIG.~\ref{fig:workflow}(e). 
It replaces the full spectral weight with a single effective mode for each particle-hole excitation.

Because in Eqn.~\eqref{eq: response to Heff}, $\mathbf{F}$ is by definition a bosonic quantity, its spectral function must satisfy 
$\boldsymbol{\rho}_A(-\Omega) = -\boldsymbol{\rho}_A(\Omega)$, \textit{i.e.} poles appear in antisymmetric pairs. 
Each diagonal element of $\mathbf{F}$ is accordingly modeled as
\begin{equation}
\begin{split}
F^{\mathrm{mod}} (z) &\approx F_{\infty} + \frac{S}{z - \Omega_p } - \frac{S}{z + \Omega_p } \\
&= F_{\infty}  +\frac{2\Omega_pS}{z^2- \Omega_p^2 }, 
\end{split}
\end{equation}
where $z$ is complex frequency,
$\Omega_p$ is pole location, and $S$ is pole strength.
The constant $F_\infty$ vanishes because 
$\mathbf{F}(i\Omega_n) \sim \mathcal{O}(1/\Omega_n)$ at large imaginary 
frequencies as
\begin{equation}
\begin{split}
    {\bf F}_\infty &= \lim_{\Omega_n \rightarrow \infty}\frac{1}{i\Omega_n \cdot {\bf I} - {\bf \Lambda}(i\Omega_n)} \\
    &= 
    \lim_{\Omega_n \rightarrow \infty}\frac{i\Omega_n \cdot {\bf I} + {\bf \Lambda}(i\Omega_n)}{- \Omega^2_n \cdot {\bf I} - [{\bf \Lambda}(i\Omega_n)]^2} \sim \mathcal{O}\left(\frac{1}{\Omega_n\cdot \mathbf{I}}\right).
\end{split}
\end{equation}

Since ${\bf \Lambda}$ and ${\bf F}$ are diagonal, each diagonal element is fitted independently. 
The two parameters $(\Omega_p, S)$ are determined by minimizing the residual
\begin{equation}
    \Delta^\mathrm{res} = \int d \Omega\,\bigg[\mathrm{Re}(F^{\mathrm{mod}} - F)\bigg]^2,
    \label{eq:delta_plasmon_pole}
\end{equation}
via least squares method.
Note that $\Delta^\mathrm{res}$ is an integral rather than a sum because of the sparsely sampled IR grid. This integral is evaluated with trapezoidal quadrature weights.

The full spectral function is then reconstructed by summing over all 
$2n_vn_c$ poles,
\begin{equation}
\label{eq:spectral_func}
    \rho^{\mathrm{tot}}_A (\Omega) = \sum_{\Omega_p}^{2n_vn_c} \frac{S}{\Omega-\Omega_p+i\eta},
\end{equation}
with a small broadening $i\eta$. 

The one-pair plasmon pole model assumes each diagonal element of $\textbf{F}$ is associated with a single excitation. 
This assumption is reasonable for weakly correlated systems because an electron-hole excitation is primarily governed by a specific occupied-virtual MO pair with minor off-diagonal coupling.

Nonetheless, approximated diagonalization and single-pole model will introduce errors, particularly at large $i\Omega_n$.
More sophisticated analytic continuation approaches could improve the dynamical correction. 
For instance, Padé approximation is commonly used to continue bosonic quantities.~\cite{vidbergSolvingEliashbergEquations1977,hanAnalyticContinuationPade2017}
More recently, Nevanlinna analytic continuation scheme has been extended to bosonic functions.~\cite{feiNevanlinnaAnalyticalContinuation2021a}
Zhang \textit{et al.} proposed a minimal-pole fitting framework for both fermionic and bosonic quantities.~\cite{zhangMinimalPoleRepresentation2024}
In addition, standalone analytic continuation of the screened Coulomb interaction $W$ has been investigated as well.~\cite{ducheminRobustAnalyticContinuationApproach2020}
We plan to pursue these directions in future work.

\subsection{Situating BSE@sc$GW$ among existing implementations}
The theoretical background above constitutes the implementation we refer to as BSE@sc$GW$ in this study. 
Several methodological choices distinguish our BSE@sc$GW$ implementation from existing approaches in the literature, each with direct consequences for the accuracy and robustness of neutral excitation energies.

The most fundamental distinction concerns the level of self-consistency in the underlying $GW$ calculation. 
The majority of BSE implementations are built on one-shot $G_0W_0$ quasiparticle energies, which introduce a well-known dependence on the choice of mean-field starting point.~\cite{brunevalBenchmarkingStartingPoints2013b,brunevalSystematicBenchmarkInitio2015} 
Our approach instead uses a fully self-consistent $GW$ reference, eliminating this ambiguity. 
Furthermore, the sc$GW$ iterations are performed entirely on the imaginary Matsubara axis using sparse sampling, in contrast to the common practice of working directly on the real frequency axis.~\cite{yehFullySelfconsistentFinitetemperature2022a}

The treatment of dynamical screening also differs from earlier work. Loos and Blase~\cite{loosDynamicalCorrectionBethe2020} introduced a dynamical correction via perturbative linearization of $W(\omega)$ around the static excitation energy, with the result renormalized by a quasiparticle weight $Z$. 
Here, instead of linearizing, we construct a bosonic response function directly from the non-Hermitian BSE Hamiltonian evaluated at each sampled Matsubara frequency. 
This provides a more direct treatment of frequency-dependent screening.

\section{Computational Details}
The experimental geometries of all molecules are taken from the Computational Chemistry Comparison and Benchmark DataBase (CCCBDB),~\cite{johnsonComputationalChemistryComparison2002a} with the exception of ethene-1,2-diaminium cation, which is not available in the CCCBDB database. 
The geometry of ethene-1,2-diaminium cation (referred to as streptocyanine-C1) is instead adopted from Ref.~\cite{loosMountaineeringStrategyExcited2018}.
Each data entry begins with a base HF mean-field calculation performed using \texttt{pyscf} version 2.8.0~\cite{sunLibcintEfficientGeneral2015b,sunPySCFPythonbasedSimulations2018a,sunRecentDevelopmentsPySCF2020a} with either cc-pVXZ or aug-cc-pVXZ basis sets.~\cite{dunningGaussianBasisSets1989,kendallElectronAffinitiesFirstrow1992,pritchardNewBasisSet2019}
In addition to the standard mean-field output, we also generate the DF-RI two-electron integral with \texttt{pyscf}.~\cite{dunlapRobustVariationalFitting2000,wernerFastLinearScaling2003,renResolutionofidentityApproachHartree2012,yeFastPeriodicGaussian2021}

The mean-field results then serve as the input for the Green’s functions in the \texttt{green-mbpt} module within the \texttt{Green/WeakCoupling} version 0.2.4.~\cite{iskakovGreenWeakCouplingImplementation2025} 
We adapt the sparse-sampled Matsubara frequency grid with intermediate representation (IR)~\cite{shinaokaCompressingGreensFunction2017a,liSparseSamplingApproach2020a} from the package \texttt{green-grids} for all sc$GW$ and BSE calculations.~\cite{iskakovGreenWeakCouplingImplementation2025}
The IR grid comprises 142 fermionic $\tau$ points and 133 bosonic $\tau$ points, using a cutoff of $\lambda = 10^5$ a.u.
All sc$GW$ calculations are conducted at the finite temperature of $\beta = 1000$ (a.u.)$^{-1}$.
Total energies calculated by sc$GW$ are converged under $10^{-7}$ a.u. 

Subsequently, BSE calculations are performed using the \texttt{green-bse} package. 
Both singlets and triplets are calculated based on the same spin-restricted sc$GW$ with different $\kappa$ values in Eqn.~\eqref{eq: kappa}.
TDA is not employed in these BSE calculations. 
The photoexcitation spectra are presented as spectral functions, as defined in Eqn.~\eqref{eq:spectral_func}. 
\texttt{molden} files used for molecular orbital visualization and excitation character assignment are generated according to the workflow described in Appendix~\ref{sec: transition type}. 
In this study, we refer to our implementation as BSE@sc$GW$.
For reference, the BSE@sc$GW$ code is archived in a dedicated branch repository.
It can be accessed on Zenodo under the name \texttt{green-bse/paper-reference-bse-scgw}.~\cite{wenGreenbsePaperreferencebsescgw2026} 
The shorthand name BSE@$G_0W_0$ used in the Results and Discussion section corresponds to the implementation and data reported by Loos and Blase in Ref.~\cite{loosDynamicalCorrectionBethe2020}.

\section{Results and discussion}

\begin{figure*}
    \centering
    \includegraphics[width=0.80\linewidth]{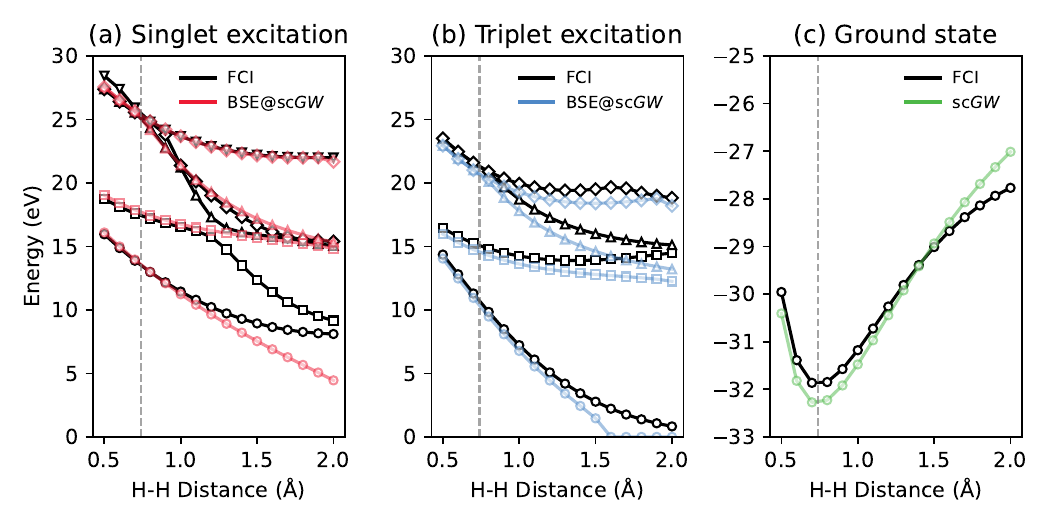}
    \caption{(a) The first four singlet excitations and (b) the first four triplet excitations of H$_2$ molecule calculated with dynamically corrected BSE@sc$GW$. (c) The ground states calculated with sc$GW$.
    Black curves are reference FCI results.    Both BSE@sc$GW$ and FCI calculations employ cc-pVTZ basis set.
    The equilibrium bond length (0.74 Å) is indicated by vertical dashed lines.}
    \label{fig:H2_exc}
\end{figure*}
\begin{figure}
    \centering
    \includegraphics[width=0.95\linewidth]{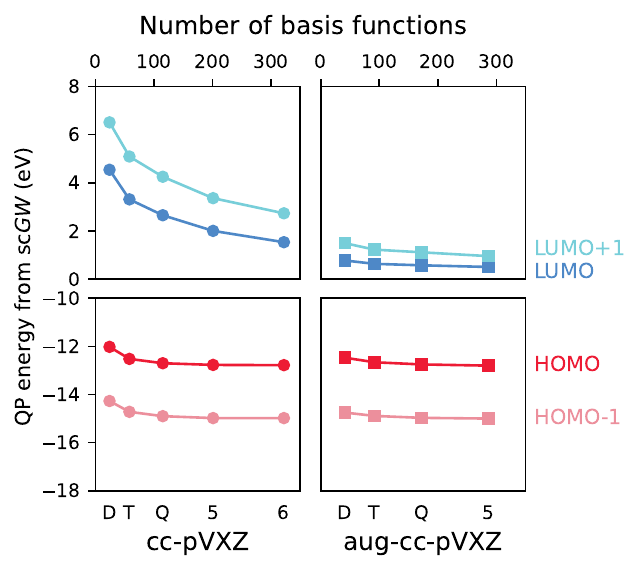}
    \caption{Convergent behavior of water QP energy levels (HOMO$-1$, HOMO, LUMO, LUMO$+1$).
    Calculated with sc$GW$ using cc-pVXZ (X = D, T, Q, 5, 6) and aug-cc-pVXZ (X = D, T, Q, 5) basis sets.}
    \label{fig:QP_ener}
\end{figure}
\begin{figure*}
    \centering
    \includegraphics[width=0.80\linewidth]{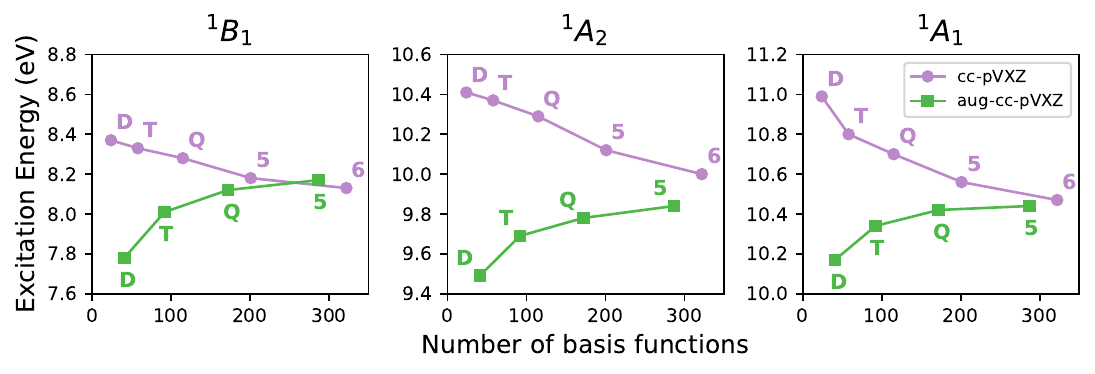}
    \caption{Convergent behavior of the first three water singlet excitations ($^1B_1$, $^1A_2$, and $^1A_1$). 
    Calculated with dynamically corrected BSE@sc$GW$, using cc-pVXZ (X = D, T, Q, 5, 6) and aug-cc-pVXZ (X = D, T, Q, 5) basis sets. 
    }
    \label{fig:basis_conv}
\end{figure*}

\subsection{\label{sec:h2-dissociation}Stretched H$_2$} 

First, to benchmark our BSE@sc$GW$ implementation on a simple case, we calculated the lowest singlet and triplet excitation energies of the H$_2$ molecule as a function of the H-H bond length using the cc-pVTZ basis set in FIG.~\ref{fig:H2_exc}.
For reference, we also performed full configuration interaction (FCI)~\cite{szaboModernQuantumChemistry1996} calculations with \texttt{pyscf} in the same basis. 

Near the equilibrium geometry (bond length = 0.74 Å), the dynamical BSE@sc$GW$ excitation energies are very accurate and closely follow the FCI reference for both singlet and triplet states. 
This agreement is expected, as near equilibrium the electronic structure of H$_2$ is well described by a single-reference picture.
The HOMO-LUMO gap remains large, and the ground state is dominated by a single Slater determinant, and the quasiparticle description underlying the $GW$ approximation is well justified. 
In this regime, the BSE Hamiltonian reliably describes the electron-hole interactions, rendering an accurate singlet-triplet splitting.

However, as the H-H distance is progressively stretched, the BSE@sc$GW$ description deteriorates and eventually breaks down in the dissociation regime.
For the singlet states, this manifests not only as a growing deviation from the FCI curve, but also as an incorrect ordering of singlet energy levels at around 1.2 Å. 
While the high-lying states still agree reasonably well with the FCI reference in this range, the lowest singlet and triplet states diverge substantially from the FCI curve approaching the dissociation limit.
Similarly, the underlying sc$GW$ reference exhibits the same trend. 
The ground-state energies obtained from sc$GW$ are consistently lower than those from FCI and show a comparable curvature in the range of 0.5 to 1.0 Å. 
However, as the interatomic separation increases, the sc$GW$ ground state energy becomes unreliable.

In addition, we observe that the triplet excitations exhibit a larger systematic deviation from FCI than the singlet excitations, resulting in a more significant offset for the triplet states over the entire bond-length range.
This offset can be understood from the structure of the BSE Hamiltonian.
The singlet-triplet splitting is calculated with different $\kappa$ values in Eqn.~\eqref{eq: kappa}.
Since the coefficient $\kappa$ is zero for $U$ for triplets, all exchange-term contributions vanish, any inaccuracy in the screened interaction $W$ in the stretched geometry becomes more pronounced. 
This issue is less important for the small molecules we examine later, since the breakdown of the underlying $GW$ reference only happens in severely non-equilibrium geometries.

This breakdown for H$_2$ at stretched bond lengths arises from strong electron correlation and the presence of nearly degenerate states. 
In such stretched geometries, the underlying $GW$ approach (and quasiparticle approximation on top) is no longer applicable for H$_2$.
As the bond is elongated, the exact ground state develops a strong multi-reference character. 
In this regime, a perturbative expansion around a single Slater determinant is no longer valid, leading to qualitative breakdowns in the description of the lowest excited states.
This also affects the correctness of the energy level ordering for the N$_2$ molecule in the following data set.
Overall, our dynamically corrected BSE@sc$GW$ performs similarly with previously reported dynamical and static BSE@$G_0W_0$ benchmark study of stretched H$_2$,~\cite{loosStaticDynamicBethe2022} confirming that the breakdown at large bond lengths is a systematic limitation of the $GW$-based approach rather than a specific deficiency of the present implementation.

\subsection{Basis set convergence}
To establish the basis set convergence of the proposed BSE method, we consider neutral excitations for the water molecule.
Specifically, we calculated the first three singlet excitations of water molecules with BSE@sc$GW$. 
Results are shown for two different basis set families: cc-pVXZ (X = D, T, Q, 5, 6) and aug-cc-pVXZ (X = D, T, Q, 5).
Larger values of $X$ correspond to basis sets with higher Dunning zeta numbers, forming a hierarchical sequence within the same basis set family.

In Fig.~\ref{fig:QP_ener}, we present the QP energy levels of HOMO$-1$, HOMO, LUMO, LUMO$+1$ calculated from the sc$GW$ reference and used as input for BSE. 
As expected, the QP energy levels exhibit a more well-behaved pattern for aug-cc-pVXZ than for cc-pVXZ, particularly for the unoccupied states. 
Virtual orbitals are inherently more spatially diffuse than occupied orbitals.
The standard cc-pVXZ basis sets, which are optimized primarily for the compact description of ground states, lack the diffuse functions necessary to represent these extended charge distributions accurately. 
As a result, the unoccupied QP levels are poorly described at lower Dunning zeta numbers ($X$) and shift significantly as additional basis functions are progressively added. 
The aug-cc-pVXZ basis sets extend the same zeta number hierarchy with extra diffuse functions. 
The resulting QP energies show a gentler and more consistent convergence pattern for both HOMO and LUMO energy levels.

Fig.~\ref{fig:basis_conv} shows how three singlet excitation energies obtained with BSE@sc$GW$ converge for the two different basis set families. 
The corresponding numerical values are provided in Table S1 of the Supplementary Material.
For the cc-pVXZ series, since the electron-hole energy differences enter as inputs in the BSE Hamiltonian, this instability in QP energy levels propagates directly into the excitation energies, leading to the poor convergence observed in Fig.~\ref{fig:basis_conv}.
By contrast, aug-cc-pVXZ basis sets better represent both the spatially extended virtual orbitals and the associated screening effects in $W$.
It leads to more stable QP gaps, and hence excitation energies converge well with respect to the number of basis functions. 

However, it should be noted that converged results in both cc-pVXZ and aug-cc-pVXZ series ultimately yield comparable excitation energies for the three singlet states once a sufficiently large number of basis functions are employed, as shown in Fig.~\ref{fig:basis_conv}. 
This convergence to a common limit confirms that the differences observed at smaller basis sets are a consequence of incomplete basis representations rather than any fundamental inconsistency between the two families. 
In general, results in the aug-cc-pVXZ basis set family produces slightly lower excitation energies, which agree more closely with the experimental values reported for the water molecule,~\cite{chutjianElectronimpactExcitationH2O1975,rubioExcitedStatesWater2008}
reflecting the improved description of the diffuse character of the excited-state wavefunctions. 
Based on these observations, aug-cc-pVXZ basis sets are used for the subsequent benchmark tests for small molecule sets in this work.

In the Supplementary Information, we also report the atomic \textit{K}-edge excitation energies of selected molecules, calculated using BSE@sc$GW$/aug-cc-pVXZ. 
Along with these results, we include a brief discussion of the basis set convergence behavior for inner-shell excitations.

\subsection{Small molecule data sets}

\begin{table*}
\caption{Singlet excitation energies of Set (a) in eV, calculated with BSE@sc$GW$/aug-cc-pVTZ.
$^\dagger$Reference values are taken from Ref.~\cite{loosDynamicalCorrectionBethe2020}. 
CCSD values were originally reported in Ref.~\cite{purvisFullCoupledclusterSingles1982}
and theoretical best estimation (TBE) values were originally reported in  Ref.~\cite{loosMountaineeringStrategyExcited2018}.}
\begin{tabular*}{0.95\textwidth}{@{\extracolsep{\fill}}cccrrrrlrr}
\hline \hline
  & & & \multicolumn{4}{c}{BSE@sc$GW$} & & \multicolumn{2}{c}{Reference$^\dagger$} \\
\cline{4-7} \cline{9-10}
Molecule & QP gap & Term symbol & $\Omega^{\mathrm{stat}}$ & $\Omega^{\mathrm{dyn}}$ & $\Delta\Omega$ & $\Delta^\mathrm{res}$ & & CCSD & TBE \\
\hline 
HCl & 12.94 & $^1\Pi$ & 8.16 & 8.12 & -0.04 & 0.0078 &  & 7.91 & 7.84 \\ \hline
\multirow{3}{*}{H$_2$O} & \multirow{3}{*}{13.29} & $^1B_1$ & 8.03 & 8.01 & -0.02 & 0.0046 &  & 7.60 & 7.17 \\ 
 &  & $^1A_2$ & 9.71 & 9.69 & -0.02 & 0.0033 &  & 9.36 & 8.92 \\
 &  & $^1A_1$ & 10.36 & 10.34 & -0.02 & 0.0033 &  & 9.96 & 9.52 \\ \hline
\multirow{7}{*}{N$_2$} & \multirow{7}{*}{18.42} & $^1\Pi_g$ & 10.17 & 10.01 & -0.16 & 0.0224 &  & 9.41 & 9.34 \\
 &  & $^1\Sigma^-_u$ & 9.98 & 9.81 & -0.17 & 0.0247 &  & 10.00 & 9.88 \\
 &  & $^1\Delta_u$ & 10.62 & 10.45 & -0.16 & 0.0213 &  & 10.44 & 10.29 \\
 &  & $^1\Sigma^+_g$ & 13.44 & 13.40 & -0.03 & 0.0034 &  & 13.15 & 12.98 \\
 &  & $^1\Pi_u$ & 13.68 & 13.64 & -0.04 & 0.0035 &  & 13.43 & 13.03 \\
 &  & $^1\Sigma^+_u$ & 13.45 & 13.41 & -0.05 & 0.0041 &  & 13.26 & 13.09 \\
 &  & $^1\Pi_u$ & 13.93 & 13.89 & -0.04 & 0.0035 &  & 13.67 & 13.46 \\ \hline
\multirow{6}{*}{CO} & \multirow{6}{*}{15.51} & $^1\Pi$ & 9.26 & 9.13 & -0.13 & 0.0217 &  & 8.59 & 8.49 \\
 &  & $^1\Sigma^-$ & 10.39 & 10.25 & -0.14 & 0.0186 &  & 9.99 & 9.92 \\
 &  & $^1\Delta$ & 10.85 & 10.71 & -0.13 & 0.0166 &  & 10.12 & 10.06 \\
 &  & $^1\Sigma^+$ & 11.41 & 11.39 & -0.02 & 0.0030 &  & 11.22 & 10.95 \\
 &  & $^1\Sigma^+$ & 11.75 & 11.71 & -0.04 & 0.0040 &  & 11.75 & 11.52 \\
 &  & $^1\Pi$ & 11.64 & 11.60 & -0.04 & 0.0043 &  & 11.96 & 11.72 \\ \hline
\multirow{2}{*}{C$_2$H$_2$} & \multirow{2}{*}{11.61} & $^1\Sigma_u^-$ & 7.27 & 7.16 & -0.10 & 0.0248 &  & 7.15 & 7.10 \\
 &  & $^1\Delta_u$ & 7.62 & 7.53 & -0.09 & 0.0210 &  & 7.48 & 7.44 \\ \hline
\multirow{3}{*}{C$_2$H$_4$} & \multirow{3}{*}{10.82} & $^1B_{3u}$ & 7.31 & 7.30 & -0.01 & 0.0040 &  & 7.42 & 7.39 \\
 &  & $^1B_{1u}$ & 7.91 & 7.86 & -0.05 & 0.0115 &  & 8.02 & 7.93 \\
 &  & $^1B_{1g}$ & 7.95 & 7.93 & -0.01 & 0.0036 &  & 8.08 & 8.08 \\ \hline
\multirow{7}{*}{CH$_2$O} & \multirow{7}{*}{11.54} & $^1A_2$ & 5.01 & 4.95 & -0.06 & 0.0288 &  & 4.01 & 3.98 \\
 &  & $^1B_2$ & 7.72 & 7.72 & 0.00 & 0.0021 &  & 7.23 & 7.23 \\
 &  & $^1B_2$ & 8.55 & 8.54 & -0.01 & 0.0024 &  & 8.12 & 8.13 \\
 &  & $^1A_1$ & 8.59 & 8.57 & -0.02 & 0.0036 &  & 8.21 & 8.23 \\
 &  & $^1A_2$ & 8.68 & 8.67 & 0.00 & 0.0016 &  & 8.65 & 8.67 \\
 &  & $^1B_1$ & 10.02 & 9.93 & -0.09 & 0.0132 &  & 9.28 & 9.22 \\
 &  & $^1A_1$ & 10.18 & 10.01 & -0.17 & 0.0226 &  & 9.67 & 9.43 \\
 \hline
\multicolumn{3}{l}{MAE (w.r.t. CCSD)}  & 0.34 & 0.30 & & & &   \\
\multicolumn{3}{l}{RMSE (w.r.t. CCSD)} & 0.42 & 0.37 & & & &   \\
\multicolumn{3}{l}{MAE (w.r.t. TBE) }  & 0.46 & 0.40 & & & & 0.15 &    \\
\multicolumn{3}{l}{RMSE (w.r.t. TBE) } & 0.54 & 0.49 & & & & 0.20 &    \\
\hline \hline
\end{tabular*}
\label{Table:singlet}
\end{table*}
\begin{table*}
\caption{Triplet excitation energies of Set (a) in eV, calculated with BSE@sc$GW$/aug-cc-pVTZ. 
$^\dagger$Reference values are taken from Ref.~\cite{loosDynamicalCorrectionBethe2020}. 
CCSD values were originally reported in Ref.~\cite{purvisFullCoupledclusterSingles1982}
and theoretical best estimation (TBE) values were originally reported in  Ref.~\cite{loosMountaineeringStrategyExcited2018}.
}
\begin{tabular*}{0.95\textwidth}{@{\extracolsep{\fill}}cccrrrrlrr}
\hline \hline
  & & & \multicolumn{4}{c}{BSE@sc$GW$} & & \multicolumn{2}{c}{Reference$^\dagger$} \\
\cline{4-7} \cline{9-10}
Molecule & QP gap & Term symbol & $\Omega^{\mathrm{stat}}$ & $\Omega^{\mathrm{dyn}}$ & $\Delta\Omega$ & $\Delta^\mathrm{res}$  & & CCSD & TBE \\
\hline 
\multirow{3}{*}{H$_2$O} & \multirow{3}{*}{13.29} & $^3B_1$ & 7.56 & 7.53 & -0.03 & 0.0074 &  & 7.20 & 6.92 \\
 &  & $^3A_2$ & 9.53 & 9.50 & -0.03 & 0.0049 &  & 9.20 & 8.91 \\
 &  & $^3A_1$ & 9.75 & 9.71 & -0.04 & 0.0063 &  & 9.49 & 9.30 \\ \hline
\multirow{4}{*}{N$_2$} & \multirow{4}{*}{18.42} & $^3\Sigma_u^+$ & 7.92 & 7.73 & -0.19 & 0.0417 &  & 7.66 & 7.70 \\
 &  & $^3\Pi_g$ & 8.43 & 8.28 & -0.15 & 0.0298 &  & 8.09 & 8.01 \\
 &  & $^3\Delta_1$ & 8.93 & 8.74 & -0.19 & 0.0331 &  & 8.91 & 8.87 \\
 &  & $^3\Sigma_u^-$ & 9.98 & 9.81 & -0.17 & 0.0247 &  & 9.83 & 9.66 \\ \hline
\multirow{5}{*}{CO} & \multirow{5}{*}{15.51} & $^3\Pi$ & 6.51 & 6.38 & -0.13 & 0.0402 &  & 6.36 & 6.28 \\
 &  & $^3\Sigma^+$ & 8.71 & 8.55 & -0.16 & 0.0296 &  & 8.34 & 8.45 \\
 &  & $^3\Delta$ & 9.53 & 9.38 & -0.15 & 0.0242 &  & 9.23 & 9.27 \\
 &  & $^3\Sigma^-_u$ & 10.39 & 10.25 & -0.14 & 0.0186 &  & 9.81 & 9.80 \\
 &  & $^3\Sigma^+_u$ & 10.67 & 10.61 & -0.06 & 0.0078 &  & 10.71 & 10.47 \\ \hline
\multirow{3}{*}{C$_2$H$_2$} & \multirow{3}{*}{11.61} & $^3\Sigma^+_u$ & 5.76 & 5.64 & -0.13 & 0.0466 &  & 5.45 & 5.53 \\
 &  & $^3\Delta_u$ & 6.55 & 6.44 & -0.12 & 0.0337 &  & 6.41 & 6.40 \\
 &  & $^3\Sigma^-_u$ & 7.27 & 7.16 & -0.10 & 0.0248 &  & 7.12 & 7.08 \\ \hline
\multirow{3}{*}{C$_2$H$_4$} & \multirow{3}{*}{10.82} & $^3B_{1u}$ & 4.76 & 4.66 & -0.10 & 0.0523 &  & 4.46 & 4.54 \\
 &  & $^3B_{3u}$ & 7.14 & 7.13 & -0.02 & 0.0051 &  & 7.29 & 7.23 \\
 &  & $^3B_{1g}$ & 7.89 & 7.87 & -0.02 & 0.0041 &  & 8.03 & 7.98 \\ \hline
\multirow{3}{*}{CH$_2$O} & \multirow{3}{*}{11.54} & $^3A_2$ & 4.26 & 4.21 & -0.05 & 0.0342 &  & 3.56 & 3.58 \\
 &  & $^3A_1$ & 6.35 & 6.21 & -0.14 & 0.0450 &  & 5.97 & 6.06 \\
 &  & $^3B_2$ & 7.46 & 7.44 & -0.01 & 0.0041 &  & 7.08 & 7.06 \\ 
\hline                            
\multicolumn{3}{l}{MAE (w.r.t. CCSD)}        & 0.28  & 0.20  &  &  &  &  &  \\
\multicolumn{3}{l}{RMSE (w.r.t. CCSD)}       & 0.32  & 0.25  &  &  &  &  &  \\
\multicolumn{3}{l}{MAE (w.r.t. TBE)}   & 0.31  & 0.23  &  &  &  & 0.10 &  \\
\multicolumn{3}{l}{RMSE (w.r.t. TBE)}  & 0.36  & 0.30  &  &  &  & 0.13 &  \\
\hline \hline
\end{tabular*}
\label{Table:triplet}
\end{table*}

\begin{table*}
\caption{Singlet and triplet excitation energies of Set (b), calculated with BSE@sc$GW$/aug-cc-pVDZ in eV.
$^\dagger$Reference CC3 values are taken from Ref.~\cite{loosDynamicalCorrectionBethe2020}, which were originally reported in Ref.~\cite{loosMountaineeringStrategyExcited2018}.}
\setchemfig{atom sep=1.8em}
\begin{tabular*}{0.95\textwidth}
{@{\extracolsep{\fill}}cccrrrrlr}
\hline \hline
\multicolumn{1}{l}{\multirow{2}{*}{}} & \multicolumn{1}{l}{} & \multicolumn{1}{l}{} & \multicolumn{4}{c}{BSE@sc$GW$} &  & \\
\cline{4-7}
Molecule & QP gap & Term symbol & $\Omega^{\mathrm{stat}}$ (eV) & $\Omega^{\mathrm{dyn}}$ (eV) & $\Delta\Omega$ (eV) & $\Delta^\mathrm{res}$ &  & CC3$^\dagger$ \\
\hline
\multirow{5}{*}{\chemfig{=[:30]-[:-30](-[:-90]H)=[:30]O}} & \multirow{5}{*}{11.01} & $^1A''$ & 4.43 & 4.37 & -0.06 & 0.0374 &  & 3.77 \\ 
 &  & $1^1A'$ & 6.42 & 6.37 & -0.05 & 0.0160 &  & 6.67 \\
 &  & $2^1A'$ & 7.60 & 7.58 & -0.02 & 0.0050 &  & 6.99 \\
 &  & $^3A''$ & 3.60 & 3.52 & -0.08 & 0.0654 &  & 3.47 \\ 
 &  & $^3A'$ & 3.85 & 3.79 & -0.06 & 0.0452 &  & 3.95 \\ \hline
\multirow{5}{*}{\chemfig{=[:30]-[:-30]=[:30]}} & \multirow{5}{*}{9.29} & $^1B_u$ & 6.01 & 5.97 & -0.03 & 0.0133 &  & 6.25 \\
 &  & $^1A_g$ & 6.25 & 6.23 & -0.02 & 0.0058 &  & 6.68 \\
 &  & $^3B_u$ & 3.55 & 3.46 & -0.08 & 0.0681 &  & 3.36 \\
 &  & $^3A_g$ & 5.43 & 5.29 & -0.15 & 0.0563 &  & 5.21 \\
 &  & $^3B_g$ & 6.16 & 6.14 & -0.02 & 0.0067 &  & 6.20 \\ \hline
\multirow{4}{*}{\chemfig{~-~}} & \multirow{4}{*}{10.31} & $^1\Sigma_u^-$ & 5.58 & 5.49 & -0.08 & 0.0316 &  & 5.44 \\
 &  & $^1\Delta_u$ & 5.81 & 5.74 & -0.08 & 0.0269 &  & 5.69 \\
 &  & $^3\Sigma_u^+$ & 4.30 & 4.19 & -0.11 & 0.0646 &  & 4.06 \\
 &  & $^3\Delta_u$ & 5.02 & 4.92 & -0.10 & 0.0432 &  & 4.86 \\ \hline
\multirow{6}{*}{\chemfig{O=[:60](-[:120]H)-(-[:-60]H)=[:60]O}} & \multirow{6}{*}{9.67} & $^1A_u$ & 3.14 & 3.11 & -0.03 & 0.0385 &  & 2.90 \\ 
 &  & $^1B_g$ & 4.80 & 4.73 & -0.07 & 0.0389 &  & 4.30 \\
 &  & $^1B_u$ & 7.79 & 7.77 & -0.02 & 0.0051 &  & 7.55 \\
 &  & $^3A_u$ & 2.47 & 2.45 & -0.02 & 0.0471 &  & 2.49 \\
 &  & $^3B_g$ & 4.10 & 4.03 & -0.07 & 0.0466 &  & 3.91 \\
 &  & $^3B_u$ & 4.83 & 4.71 & -0.11 & 0.0596 &  & 5.20 \\ \hline
\chemfig{H_2N-[:30]=[:-30]NH_2^{\oplus}} & 13.23 & $^1B_2$ & 7.47 & 7.43 & -0.05 & 0.0131 &  & 7.14 \\
\hline
\multicolumn{3}{l}{MAE (w.r.t. CC3)} & 0.26 & 0.23 &  &  &  & \\
\multicolumn{3}{l}{RMSE (w.r.t. CC3)} & 0.31 & 0.29 &  & &  & \\
\hline
\hline
\end{tabular*}
\label{Table:medium_mols}
\end{table*}

In Tables ~\ref{Table:singlet}, ~\ref{Table:triplet} and ~\ref{Table:medium_mols},
we present a comprehensive data set of neutral singlet and triplet excitation energies, calculated with BSE@sc$GW$. 
The data set lists the lowest singlet and triplet excitations for two sets of molecules.
Set (a) comprises seven small molecules (HCl, H$_2$O, N$_2$, CO, C$_2$H$_2$, C$_2$H$_4$, and CH$_2$O). 
These molecules account for 29 singlet and 21 triplet excitations reported in Table~\ref{Table:singlet} and Table~\ref{Table:triplet}, respectively.
Set (b) comprises five medium sized molecules with three or four non-hydrogen atoms (acrolein, butadiene, diacetylene, glyoxal, and streptocyanine-C1).
These medium molecules account for 11 singlet and 10 triplet excitations listed in Table~\ref{Table:medium_mols}.

For each excitation, we report both the dynamically corrected solution $\Omega^{\mathrm{dyn}}$, the static solution $\Omega^{\mathrm{stat}}$. 
Along with the excitation energies, we also present the dynamical corrections defined as $\Delta \Omega = \Omega^{\mathrm{dyn}} - \Omega^{\mathrm{stat}}$, and the residual error $\Delta^{\mathrm{res}}$ in plasmon-pole approximation defined in Eqn.~\eqref{eq:delta_plasmon_pole}.
The residual serves as a diagnostic indicator of the accumulated difference between the fitted results and the original sparsely sampled auxiliary function $\textbf{F}(i\Omega_n)$.
Note the residual $\Delta^{\mathrm{res}}$ solely reflects the fitting quality underlying the dynamical treatment, it is not a measure of energy.

For reference, in Set (a) we compare BSE@sc$GW$ against BSE@$G_0W_0$@HF, CCSD,~\cite{purvisFullCoupledclusterSingles1982} and theoretical best estimation (TBE) data reported by Loos and Blase.~\cite{loosDynamicalCorrectionBethe2020} 
The TBE values were calculated at the exFCI/aug-cc-pVTZ level, and basis set corrections were applied to selected entries.~\cite{loosMountaineeringStrategyExcited2018}
In Set (b), we compare BSE@sc$GW$ against BSE@$G_0W_0$@HF and CC3~\cite{kochCC3ModelIterative1997} results.~\cite{loosDynamicalCorrectionBethe2020}


BSE@sc$GW$ achieves accuracy comparable with wave function-based approaches, including CCSD and CC3.
For the singlet states of Set (a), the static BSE@sc$GW$ yields mean absolute error (MAE) and root mean square error (RMSE) are 0.34 and 0.42 eV, respectively with respect to CCSD; these values decrease to 0.30 and 0.37 eV when the dynamical correction is taken into account. 
For the triplet states of Set (a), the MAE and RMSE drop from 0.28/0.32 eV to 0.20/0.25 eV. 
A comparison between BSE@sc$GW$ and the TBE indicates similar trends.
Although it does not reach the absolute accuracy of CCSD, the dynamically corrected BSE@sc$GW$ achieves a systematic reduction of the error. 
This behavior aligns with the moderate yet physically significant influence of frequency-dependent screening. 
For the singlets and triplets of medium-sized molecules in Set (b), the static BSE@sc$GW$ yields a MAE/RMSE of 0.26/0.31 eV with respect to CC3, which is reduced to 0.23/0.29 eV upon inclusion of the dynamical correction. 
The improvement is in line with the trend observed for Set (a).

\begin{table}
\caption{Errors of Set~(a) in eV. BSE@sc$GW$ and BSE@$G_0W_0$@HF are compared against the CCSD and TBE benchmarks. Singlets (top) and triplets (bottom) are listed separately. 
$^\dagger$Reported by (or derived from data presented by) Loos and Blase.~\cite{loosDynamicalCorrectionBethe2020}} 
\begin{tabular*}{0.95\columnwidth}{@{\extracolsep{\fill}}ccrrlrr}
\hline
\hline 
\multicolumn{2}{c}{Singlets} & \multicolumn{2}{c}{BSE@sc$GW$} &  & \multicolumn{2}{c}{BSE@$G_0W_0$@HF$^\dagger$} \\
\cline{3-4} \cline{6-7}
Reference & Error & $\Omega^{\mathrm{stat}}$ & $\Omega^{\mathrm{dyn}}$ &  & $\Omega^{\mathrm{stat}}$ & $\Omega^{\mathrm{dyn}}$  \\
\hline
\multirow{2}{*}{CCSD} & MAE & 0.34 & 0.30 &  & 0.50 & 0.38 \\
 & RMSE & 0.42 & 0.37 &  & 0.56 & 0.43 \\
 \hline
\multirow{2}{*}{TBE} & MAE & 0.46 & 0.40 &  & 0.64 & 0.50 \\
 & RMSE & 0.54 & 0.49 &  & 0.70 & 0.58 \\
\hline
\hline
\multicolumn{2}{c}{Triplets} & \multicolumn{2}{c}{BSE@sc$GW$} &  & \multicolumn{2}{c}{BSE@$G_0W_0$@HF$^\dagger$} \\
\cline{3-4} \cline{6-7}
Reference & Error & $\Omega^{\mathrm{stat}}$ & $\Omega^{\mathrm{dyn}}$ &  & $\Omega^{\mathrm{stat}}$ & $\Omega^{\mathrm{dyn}}$ \\
\hline
\multirow{2}{*}{CCSD} & MAE & 0.28 & 0.20 &  & 0.36 & 0.21 \\
 & RMSE & 0.32 & 0.25 &  & 0.39 & 0.25 \\
 \hline
\multirow{2}{*}{TBE} & MAE & 0.31 & 0.23 &  & 0.41 & 0.27 \\
 & RMSE & 0.36 & 0.30 &  & 0.45 & 0.33 \\
\hline
\hline 
\end{tabular*}
\label{Table:statistic_a}
\end{table}
\begin{table}
\caption{Errors of Set~(b) in eV. BSE@sc$GW$ and BSE@$G_0W_0$@HF are compared against the CC3 benchmarks.
$^\dagger$Reported by (or derived from data presented by) Loos and Blase.~\cite{loosDynamicalCorrectionBethe2020}}
\begin{tabular*}{0.95\columnwidth}{@{\extracolsep{\fill}}cclrllr}
\hline
\hline
 & & \multicolumn{2}{c}{BSE@sc$GW$} &  & \multicolumn{2}{c}{BSE@$G_0W_0$@HF$^\dagger$} \\
\cline{3-4} \cline{6-7}
Reference & Error & $\Omega^{\mathrm{stat}}$ & $\Omega^{\mathrm{dyn}}$ &  & $\Omega^{\mathrm{stat}}$ & $\Omega^{\mathrm{dyn}}$ \\
\hline
\multirow{2}{*}{CC3} & MAE & \multicolumn{1}{r}{0.26} & 0.23 &  & \multicolumn{1}{r}{0.32} & 0.23 \\
 & RMSE & \multicolumn{1}{r}{0.31} & 0.29 &  & \multicolumn{1}{r}{0.38} & 0.29 \\
\hline
\hline
\end{tabular*}
\label{Table:statistic_b}
\end{table}

In Tables~\ref{Table:statistic_a} and \ref{Table:statistic_b}, we aggregate the MAEs and RMSEs of BSE@sc$GW$ and BSE@$G_0W_0$@HF of Loos and Blase~\cite{loosDynamicalCorrectionBethe2020} with respect to the same referential values.  
The static BSE@sc$GW$ consistently outperforms the static BSE@$G_0W_0$ for both sets. 
In some cases, the static BSE@sc$GW$ even gives lower errors than dynamical BSE@$G_0W_0$. 
For instance, in singlets of Set (a), BSE@sc$GW$ $\Omega^{\mathrm{stat}}$ has a MAE/RMSE of 0.34/0.42 eV (w.r.t.~CCSD) and 0.46/0.54 eV (w.r.t.~TBE), while BSE@$G_0W_0$ $\Omega^{\mathrm{dyn}}$ has larger 0.38/0.43 eV and 0.50/0.58 eV respectively, despite with dynamical effects included.
For both BSE@sc$GW$ and BSE@$G_0W_0$, the dynamically corrected results performed about the same for this test set. 
The benefit of introducing dynamical corrections to the BSE kernel was found to be limited in magnitude for BSE@sc$GW$.
The dynamical correction was observed to be more significant for the less accurate static BSE@$G_0W_0$ calculations, where it leads to a more noticeable improvement. 
In contrast, BSE@sc$GW$, which is already more reliable due to self-consistency, shows a comparatively smaller but still systematic benefit from the dynamical treatment.
Nevertheless, once the dynamical correction was applied, the results obtained with BSE@sc$GW$ and BSE@$G_0W_0$ became more similar to each other than in the static case. 
This trend suggests that the dynamical correction is physically meaningful for BSE@sc$GW$, even if its numerical impact on the excitation energies is only moderately beneficial.

We would like to point out one peculiar observation for the excitation spectra of N$_2$, noted in Table.~\ref{Table:singlet}. 
The wave-function-based CCSD and TBE reference values predict $^1\Sigma_u^-$ to be the lowest-lying singlet excitation. 
But for BSE@sc$GW$, the ordering of $^1\Sigma_u^-$ and $^1\Pi_g$ is wrongly predicted. 
This behavior is also observed in BSE@$G_0W_0$ by Loos and Blase,~\cite{loosStaticDynamicBethe2022} and echoes the quasiparticle picture breakdown discussed in the benchmark for stretched H$_2$ molecule (cf., Sec.~\ref{sec:h2-dissociation}). 
Furthermore, the $GW$ approximation is perhaps insufficient for the N$_2$ molecule, given its multi-reference character and the presence of closely spaced energy levels.~\cite{aryasetiawanGWMethod1998}

For both data sets (a) and (b) combined, dynamical BSE@sc$GW$ provides an average of 0.06 eV of correction upon the static results for singlets, and 0.09 eV for triplets. 
This is consistent with the observation of Rohlfing \textit{et al.} that dynamical correction only accounts for minimal changes of about 0.1 eV for valence shell excitations.~\cite{rohlfingElectronholeExcitationsOptical2000}

\section{Conclusions}

The BSE@sc$GW$ approach yields accurate neutral excitation energies for small molecules.
At the static limit, it systematically outperforms existing reported BSE@$G_0W_0$ results,~\cite{loosDynamicalCorrectionBethe2020}
for molecules in Sets~(a) and~(b).
The dynamically corrected BSE@sc$GW$ likewise surpasses the dynamically corrected BSE@$G_0W_0$ in Set~(a).
For the medium-sized molecule in Set~(b), the performance of dynamical BSE@sc$GW$ and dynamical BSE@$G_0W_0$ is comparable.
Since both the static and dynamic BSE@sc$GW$ results agree closely with the reference values, we conclude that the self-consistent $GW$ scheme is well suited to serve as the reference state for BSE calculations.
In general, although the dynamical correction in our scheme is smaller in magnitude compared to BSE@$G_0W_0$, it still leads to a clear improvement over the static calculations.
Despite substantial differences in how dynamical corrections are implemented in practice, their frequency dependence has a common origin: the screened Coulomb interaction, which introduces non-linearity into the interaction kernel $\Xi$.

To solve the BSE in the non-relativistic Casida formalism, we adopt a series of controlled approximations.
We invoke the quasiparticle approximation within sc$GW$ to generate reliable input for the BSE interaction kernel.
Addressing the frequency dependence of the interaction kernel $\Xi$ requires additional approximations specific to our dynamical scheme.
We adopt an adiabatic approximation, in which eigenstates corresponding to distinct electron-hole excitations do not mix at non-zero frequency.
Consequently, the auxiliary response function $\textbf{F}$ can be expressed in terms of a fixed set of particle-hole eigenstates, and the dynamical coupling effect between excitations is neglected.
The excitation manifold therefore remains diagonal in the occupied-virtual MO basis at all frequencies, and the bosonic frequency dependence of the electron-hole excitations is treated with the plasmon-pole model.
Together, these approximations simplify the structure of the response function and establish a direct mapping between individual excitations and their spectral functions.

Our BSE@sc$GW$ implementation differs from standard approaches in three respects: 
it uses a fully self-consistent $GW$ reference rather than $G_0W_0$, eliminating starting-point dependence; 
all calculations are performed on the imaginary time and frequency axes; 
and dynamical screening is treated by constructing a bosonic response function directly from the frequency-dependent BSE Hamiltonian.

A key limitation of this framework is its inability to describe states with pronounced multi-reference character, such as double excitations, molecules with stretched geometries, and coupled particle-hole eigenstates. 
In order to address this issue, it is necessary to go beyond the quasiparticle approximation.
Treating such multi-reference states will require either a sc$GW$ scheme that retains off-diagonal self-energy contributions or embedding strategies that incorporate strong correlation in a localized subspace.

In the future, we intend to further develop BSE@sc$GW$ in several directions.
First, the current implementation can be refined both at the numerical and theoretical levels.
Instead of fully diagonalizing the effective Hamiltonian, which scales poorly with the size of the occupied-virtual MO space, one could employ advanced iterative eigensolvers.
For example, approaches like the Davidson algorithm employ physically motivated initial guess vectors to quickly converge to the lowest few eigenstates, thereby lowering both memory usage and computational cost.
More advanced analytical continuation methods, such as Padé and Nevanlinna, can be adapted to render dynamical BSE@sc$GW$ results instead of the current crude plasmon-pole model.

\section*{Supplementary material}
See Supplementary Material for: 
(i) Geometries of all molecules in Sets (a) and (b);
(ii) Underlying data used to plot Figure~\ref{fig:basis_conv};
(iii) A supplemental discussion of \textit{K}-edge excitations;
(iv) Archived raw BSE@sc$GW$ output logs for Sets (a) and (b). 

\section*{Acknowledgements}
The author would like to thank Lei Zhang for insightful discussions on analytic continuation techniques for bosonic functions. 
This study is supported by the U.S. Department of Energy, Office of Science, Office of Advanced Scientific Computing Research and Office of Basic Energy Sciences, Scientific Discovery through Advanced Computing (SciDAC) program under Award No.~DE-SC0022198. 
M.W. is also supported by the National Science Foundation (NSF) through the Materials Research Science and Engineering Center (MRSEC) at the University of Michigan under Award No.~DMR-2309029.

\section*{Author declarations}
\subsection*{Conflict of interest}
The authors have no conflicts of interest to disclose.
\subsection*{Author Contributions}
\textbf{Ming Wen}: 
Conceptualization (supporting);
Formal analysis (equal);
Investigation (lead);
Methodology (lead); 
Data curation (lead); 
Software development (lead); 
Visualization (lead);
Original draft (lead);
Review \& editing (supporting).
\textbf{Gaurav Harsha}:
Conceptualization (supporting);
Formal analysis (equal);
Software development (supporting); 
Original draft (supporting);
Review \& editing (equal).
\textbf{Dominika Zgid}: Conceptualization (lead); 
Funding acquisition (lead);
Resources (lead);
Project administration (lead);
Supervision (lead);
Review \& editing (equal).

\section*{Data availability}

A dedicated reference branch repository \texttt{green-bse}/\texttt{paper-reference-bse-scgw} used in this work is available on Zenodo.~\cite{wenGreenbsePaperreferencebsescgw2026} 

Further data supporting the results of this study are available from the corresponding author upon reasonable request.

\appendix

\section{Wick rotation}
\label{sec: wick}

The Matsubara GF defined in Eqn.~\eqref{eq:G_tau_definition} is related to its real-time counterpart through a Wick rotation.~\cite{fetterQuantumTheoryMany1971}
The real-time GF is defined as
\begin{equation}
    G_{pq}(t) = -\frac{i}{\mathcal{Z}} \mathrm{Tr} \left[
        e^{-\beta(H-\mu N)} T_t \left( c_p(t) c_q^\dagger(0) \right)
    \right],
\end{equation}
where $c_p(t) = e^{iHt} c_p e^{-iHt}$ is the time-dependent annihilation operator, $T_t$ is the time-ordering operator in the real-time Heisenberg picture. 
The Wick rotation is the substitution of $t \to -i\tau$.
The unitary time-evolution operator becomes a decaying exponential:
\begin{equation}
    e^{-iHt} \to e^{-\tau H},
\end{equation}
which rotates the time contour from the real axis to the imaginary axis in the complex time plane.
The imaginary-time Heisenberg-picture annihilation operator becomes
\begin{equation}
    c_p(t)\big|_{t=-i\tau} = e^{\tau H} c_p e^{-\tau H} \equiv c_p(\tau),
\end{equation}
Simultaneously, the prefactor transforms from $-i$ to $-1$, recovering the sign convention in Eqns.~\eqref{eq:Pi_AO} to \eqref{eq:Sigma_AO} without the imaginary unit. 
The relation between the real-time and Matsubara Green’s functions is an analytic continuation defined on a complex-time contour:
\begin{equation}
G(t)\big|_{t=-i\tau} \;\xrightarrow\; G(\tau).
\end{equation}

\section{Determination of transition type}
\label{sec: transition type}

The eigenvector matrix $\textbf{V}$ solved from the effective Hamiltonian $\textbf{H}^{\mathrm{stat}}$ in Eqn.~\eqref{eq: static eig eq} has this block structure as
\begin{equation}
    \mathbf{V} = \begin{pmatrix}
    \textbf{X} \\
    \textbf{Y}
    \end{pmatrix}.
\end{equation}
It can be transformed from the occupied-virtual MO basis to AO basis. 
We first dissect the MO coefficient matrix into the occupied MO and virtual MO parts. 
\begin{subequations}
\begin{align}
    \textbf{C}_{v} &= \textbf{C}[:, 0:n_{v}], \\
    \textbf{C}_{c} &= \textbf{C}[:, n_{v}:n_{\mathrm{MO}}].
\end{align}
\end{subequations}
We define the mapping matrix from occupied-virtual MOs to AOs, which consists of two blocks.
The blocks $\textbf{M}_{v}$ and $\textbf{M}_{c}$ each have dimensions $(n_{\mathrm{AO}} \times 2 n_{v} n_{c})$ as
\begin{equation}  
\textbf{M} = \begin{pmatrix}
\underbrace{\textbf{M}_{b=0}}_{n_{v} n_{c}} & \underbrace{\textbf{M}_{b=1}}_{n_{v} n_{c}}
\end{pmatrix} \bigg\}{\,}_{n_{\mathrm{AO}}} .
\end{equation}
For each occupied orbital \(i\), virtual orbital \(a\), and block \(b \in \{0,1\}\), the new column index is calculated via $\alpha(i,a,b) = b \cdot n_{v} n_{c} + i \cdot n_{c} + a $.
This maps all the electron-hole excitations $(i \to a)$ to every column in both $\mathbf{M}_{v}$ and $\mathbf{M}_{c}$ as
\begin{subequations} 
\begin{align}
\textbf{M}_{v}[:, \alpha(i,a,b)] &= (-1)^b \, \textbf{C}_v[:, i], \\
\textbf{M}_c[:, \alpha(i,a,b)] &= (-1)^b \, \textbf{C}_c[:, a].
\end{align}
\end{subequations}
The rows of $\textbf{M}$ correspond to AO indices.  
The columns correspond to signed excitations.
The sign $(-1)^b$ encodes the positive and negative magnitudes for $\textbf{X}$ and $\textbf{Y}$ blocks.
We use the mapping matrices to transform the eigenvector matrix $\textbf{V}$ solved from the effective Hamiltonian $\textbf{H}^{\mathrm{stat}}$ in Eqn.~\eqref{eq: static eig eq} to AO basis as
\begin{subequations}
\begin{align}
\textbf{V}_{v} &= \textbf{M}_{v} \, \textbf{V}, \\
\textbf{V}_{c} &= \textbf{M}_{c} \, \textbf{V}.
\end{align}
\end{subequations}
$\textbf{V}_{v}$ and $\textbf{V}_{c}$, each of dimensions $(n_{\mathrm{AO}} \times n_{\mathrm{exc}})$, are the AO basis projections.
They capture the transition density $(X_{ia} - Y_{ia})$ projected onto the occupied and virtual subspaces. 
The $k$-th columns of $\textbf{V}_v$ and $\textbf{V}_c$ represent, respectively, the coefficient vectors of the occupied MO and the virtual MO that participate in the $k$-th excitation.
These two vectors are stored in a \texttt{molden} file. It can then be visualized to conveniently determine the nature of the excitation, as showcased in Figure \ref{fig:workflow} (f).

\bibliography{BSE}

@article{albrechtExcitonicEffectsOptical1998,
  title = {Excitonic {{Effects}} in the {{Optical Properties}}},
  author = {Albrecht, S. and Reining, L. and Del Sole, R. and Onida, G.},
  year = 1998,
  journal = {physica status solidi (a)},
  volume = {170},
  number = {2},
  pages = {189--197},
  issn = {1521-396X},
  doi = {10.1002/(SICI)1521-396X(199812)170:2<189::AID-PSSA189>3.0.CO;2-3},
  urldate = {2026-03-02},
  copyright = {\copyright{} 1998 WILEY-VCH Verlag Berlin GmbH, Fed. Rep. of Germany},
}

@article{aryasetiawanGWMethod1998,
  title = {The \emph{GW} Method},
  author = {Aryasetiawan, F. and Gunnarsson, O.},
  year = 1998,
  month = mar,
  journal = {Rep. Prog. Phys.},
  volume = {61},
  number = {3},
  pages = {237},
  issn = {0034-4885},
  doi = {10.1088/0034-4885/61/3/002},
  urldate = {2023-08-01},
}

@article{authierDynamicalKernelsOptical2020,
  title = {Dynamical Kernels for Optical Excitations},
  author = {Authier, Juliette and Loos, Pierre-Fran{\c c}ois},
  year = 2020,
  month = nov,
  journal = {J. Chem. Phys.},
  volume = {153},
  number = {18},
  pages = {184105},
  issn = {0021-9606},
  doi = {10.1063/5.0028040},
  urldate = {2026-02-02},
}

@article{bartlettCoupledclusterTheoryQuantum2007a,
  title = {Coupled-Cluster Theory in Quantum Chemistry},
  author = {Bartlett, Rodney J. and Musia{\l}, Monika},
  year = 2007,
  month = feb,
  journal = {Rev. Mod. Phys.},
  volume = {79},
  number = {1},
  pages = {291--352},
  publisher = {American Physical Society},
  doi = {10.1103/RevModPhys.79.291},
  urldate = {2026-01-28},
}

@article{baymConservationLawsCorrelation1961,
  title = {Conservation {{Laws}} and {{Correlation Functions}}},
  author = {Baym, Gordon and Kadanoff, Leo P.},
  year = 1961,
  month = oct,
  journal = {Phys. Rev.},
  volume = {124},
  number = {2},
  pages = {287--299},
  publisher = {American Physical Society},
  doi = {10.1103/PhysRev.124.287},
  urldate = {2026-01-27},
}

@article{baymSelfConsistentApproximationsManyBody1962,
  title = {Self-{{Consistent Approximations}} in {{Many-Body Systems}}},
  author = {Baym, Gordon},
  year = 1962,
  month = aug,
  journal = {Phys. Rev.},
  volume = {127},
  number = {4},
  pages = {1391--1401},
  publisher = {American Physical Society},
  doi = {10.1103/PhysRev.127.1391},
  urldate = {2026-01-27},
}

@article{bechstedtCompensationDynamicalQuasiparticle1997,
  title = {Compensation of {{Dynamical Quasiparticle}} and {{Vertex Corrections}} in {{Optical Spectra}}},
  author = {Bechstedt, F. and Tenelsen, K. and Adolph, B. and Del Sole, R.},
  year = 1997,
  month = feb,
  journal = {Phys. Rev. Lett.},
  volume = {78},
  number = {8},
  pages = {1528--1531},
  issn = {0031-9007, 1079-7114},
  doi = {10.1103/PhysRevLett.78.1528},
  urldate = {2026-03-03},
  copyright = {http://link.aps.org/licenses/aps-default-license},
}

@article{bintrimFullfrequencyDynamicalBethe2022,
  title = {Full-Frequency Dynamical {{Bethe}}--{{Salpeter}} Equation without Frequency and a Study of Double Excitations},
  author = {Bintrim, Sylvia J. and Berkelbach, Timothy C.},
  year = 2022,
  month = jan,
  journal = {J. Chem. Phys.},
  volume = {156},
  number = {4},
  pages = {044114},
  issn = {0021-9606},
  doi = {10.1063/5.0074434},
  urldate = {2026-01-26},
}

@article{blaseBetheSalpeterEquation2018,
  title = {The {{Bethe}}--{{Salpeter}} Equation in Chemistry: Relations with {{TD-DFT}}, Applications and Challenges},
  shorttitle = {The {{Bethe}}--{{Salpeter}} Equation in Chemistry},
  author = {Blase, Xavier and Duchemin, Ivan and Jacquemin, Denis},
  year = 2018,
  month = feb,
  journal = {Chem. Soc. Rev.},
  volume = {47},
  number = {3},
  pages = {1022--1043},
  publisher = {The Royal Society of Chemistry},
  issn = {1460-4744},
  doi = {10.1039/C7CS00049A},
  urldate = {2026-04-08},
}

@article{blaseBetheSalpeterEquation2020,
  title = {The {{Bethe}}--{{Salpeter Equation Formalism}}: {{From Physics}} to {{Chemistry}}},
  shorttitle = {The {{Bethe}}--{{Salpeter Equation Formalism}}},
  author = {Blase, Xavier and Duchemin, Ivan and Jacquemin, Denis and Loos, Pierre-Fran{\c c}ois},
  year = 2020,
  month = sep,
  journal = {J. Phys. Chem. Lett.},
  volume = {11},
  number = {17},
  pages = {7371--7382},
  issn = {1948-7185, 1948-7185},
  doi = {10.1021/acs.jpclett.0c01875},
  urldate = {2026-01-26},
  copyright = {https://doi.org/10.15223/policy-029},
}

@book{bruusManyBodyQuantumTheory2004,
  title = {Many-{{Body Quantum Theory}} in {{Condensed Matter Physics}}: {{An Introduction}}},
  shorttitle = {Many-{{Body Quantum Theory}} in {{Condensed Matter Physics}}},
  author = {Bruus, Henrik and Flensberg, Karsten},
  year = 2004,
  month = nov,
  series = {Oxford {{Graduate Texts}}},
  publisher = {Oxford University Press},
  address = {Oxford, New York},
  isbn = {978-0-19-856633-5},
}

@article{carusoBenchmarkGWApproaches2016,
  title = {Benchmark of {{GW Approaches}} for the {{GW100 Test Set}}},
  author = {Caruso, Fabio and Dauth, Matthias and {van Setten}, Michiel J. and Rinke, Patrick},
  year = 2016,
  month = oct,
  journal = {J. Chem. Theory Comput.},
  volume = {12},
  number = {10},
  pages = {5076--5087},
  publisher = {American Chemical Society},
  issn = {1549-9618},
  doi = {10.1021/acs.jctc.6b00774},
  urldate = {2023-03-01},
}

@article{casanova-paezCoreExcitedStatesOpenShell2025,
  title = {Core-{{Excited States}} for {{Open-Shell Systems}} in {{Similarity-Transformed Equation-of-Motion Theory}}},
  author = {{Casanova-P{\'a}ez}, Marcos and Neese, Frank},
  year = 2025,
  month = feb,
  journal = {J. Chem. Theory Comput.},
  volume = {21},
  number = {3},
  pages = {1306--1321},
  publisher = {American Chemical Society},
  issn = {1549-9618},
  doi = {10.1021/acs.jctc.4c01181},
  urldate = {2026-01-28},
}

@incollection{CASIDA1996391,
  title = {Time-Dependent Density Functional Response Theory of Molecular Systems: {{Theory}}, Computational Methods, and Functionals},
  booktitle = {Recent Developments and Applications of Modern Density Functional Theory},
  author = {Casida, Mark E.},
  editor = {Seminario, J.M.},
  year = 1996,
  series = {Theoretical and Computational Chemistry},
  volume = {4},
  pages = {391--439},
  publisher = {Elsevier},
  issn = {1380-7323},
  doi = {10.1016/S1380-7323(96)80093-8}
}

@article{choSimplifiedGWBSE2022,
  title = {Simplified {{GW}}/{{BSE Approach}} for {{Charged}} and {{Neutral Excitation Energies}} of {{Large Molecules}} and {{Nanomaterials}}},
  author = {Cho, Yeongsu and Bintrim, Sylvia J. and Berkelbach, Timothy C.},
  year = 2022,
  month = jun,
  journal = {J. Chem. Theory Comput.},
  volume = {18},
  number = {6},
  pages = {3438--3446},
  publisher = {American Chemical Society},
  issn = {1549-9618},
  doi = {10.1021/acs.jctc.2c00087},
  urldate = {2026-03-24},
}

@article{chutjianElectronimpactExcitationH2O1975,
  title = {Electron-impact Excitation of {{H2O}} and {{D2O}} at Various Scattering Angles and Impact Energies in the Energy-loss Range 4.2--12 {{eV}}},
  author = {Chutjian, A. and Hall, R. I. and Trajmar, S.},
  year = 1975,
  month = jul,
  journal = {J. Chem. Phys.},
  volume = {63},
  number = {2},
  pages = {892--898},
  issn = {0021-9606},
  doi = {10.1063/1.431370},
  urldate = {2026-03-30},
}

@article{cremerConfigurationInteractionCoupled2013,
  title = {From Configuration Interaction to Coupled Cluster Theory: {{The}} Quadratic Configuration Interaction Approach},
  shorttitle = {From Configuration Interaction to Coupled Cluster Theory},
  author = {Cremer, Dieter},
  year = 2013,
  journal = {WIREs Comput. Mol. Sci.},
  volume = {3},
  number = {5},
  pages = {482--503},
  issn = {1759-0884},
  doi = {10.1002/wcms.1131},
  urldate = {2026-03-24},
  copyright = {Copyright \copyright{} 2013 John Wiley \& Sons, Ltd.},
}

@article{delgrandeHowChooseEfficiently2025,
  title = {How to Choose Efficiently the Size of the {{Bethe-Salpeter}} Equation {{Hamiltonian}} for Accurate Exciton Calculations on Supercells},
  author = {Del Grande, Rafael R. and Strubbe, David A.},
  year = 2025,
  month = oct,
  journal = {Phys. Rev. B},
  volume = {112},
  number = {16},
  pages = {165118},
  publisher = {American Physical Society},
  doi = {10.1103/dg13-y4kj},
  urldate = {2026-03-24},
}

@article{ducheminRobustAnalyticContinuationApproach2020,
  title = {Robust {{Analytic-Continuation Approach}} to {{Many-Body GW Calculations}}},
  author = {Duchemin, Ivan and Blase, Xavier},
  year = 2020,
  month = mar,
  journal = {J. Chem. Theory Comput.},
  volume = {16},
  number = {3},
  pages = {1742--1756},
  publisher = {American Chemical Society},
  issn = {1549-9618},
  doi = {10.1021/acs.jctc.9b01235},
  urldate = {2026-03-26},
}

@article{dunlapRobustVariationalFitting2000,
  title = {Robust and Variational Fitting},
  author = {Dunlap, Brett I.},
  year = 2000,
  month = jan,
  journal = {Phys. Chem. Chem. Phys.},
  volume = {2},
  number = {10},
  pages = {2113--2116},
  publisher = {The Royal Society of Chemistry},
  issn = {1463-9084},
  doi = {10.1039/B000027M},
  urldate = {2023-11-09},
}

@article{dunningGaussianBasisSets1989,
  title = {Gaussian Basis Sets for Use in Correlated Molecular Calculations. {{I}}. {{The}} Atoms Boron through Neon and Hydrogen},
  author = {Dunning, Jr., Thom H.},
  year = 1989,
  month = jan,
  journal = {J. Chem. Phys.},
  volume = {90},
  number = {2},
  pages = {1007--1023},
  issn = {0021-9606},
  doi = {10.1063/1.456153},
  urldate = {2026-01-27},
}

@article{feiNevanlinnaAnalyticalContinuation2021a,
  title = {Nevanlinna {{Analytical Continuation}}},
  author = {Fei, Jiani and Yeh, Chia-Nan and Gull, Emanuel},
  year = 2021,
  month = feb,
  journal = {Phys. Rev. Lett.},
  volume = {126},
  number = {5},
  pages = {056402},
  doi = {10.1103/PhysRevLett.126.056402},
  urldate = {2023-02-27},
}

@book{fetterQuantumTheoryMany1971,
  title = {Quantum Theory of Many-Particle Systems},
  author = {Fetter, Alexander L. and Walecka, John Dirk},
  year = {1971},
  publisher = {McGraw-Hill},
  address = {New York},
}

@article{forsterQuasiparticleSelfConsistentGWBetheSalpeter2022,
  title = {Quasiparticle {{Self-Consistent GW-Bethe}}--{{Salpeter Equation Calculations}} for {{Large Chromophoric Systems}}},
  author = {F{\"o}rster, Arno and Visscher, Lucas},
  year = 2022,
  month = nov,
  journal = {J. Chem. Theory Comput.},
  volume = {18},
  number = {11},
  pages = {6779--6793},
  publisher = {American Chemical Society},
  issn = {1549-9618},
  doi = {10.1021/acs.jctc.2c00531},
  urldate = {2026-03-24},
}

@article{gantOptimallyTunedStarting2022,
  title = {Optimally Tuned Starting Point for Single-Shot \emph{GW} Calculations of Solids},
  author = {Gant, Stephen E. and Haber, Jonah B. and Filip, Marina R. and Sagredo, Francisca and Wing, Dahvyd and Ohad, Guy and Kronik, Leeor and Neaton, Jeffrey B.},
  year = 2022,
  month = may,
  journal = {Phys. Rev. Mater.},
  volume = {6},
  number = {5},
  pages = {053802},
  publisher = {American Physical Society},
  doi = {10.1103/PhysRevMaterials.6.053802},
  urldate = {2026-02-02},
}

@article{garcia-gonzalezManyBody$mathitGW$Calculations2002,
  title = {Many-{{Body}} \emph{GW} {{Calculations}} of {{Ground-State Properties}}: {{Quasi-2D Electron Systems}} and van Der {{Waals Forces}}},
  shorttitle = {Many-{{Body}} \$\textbackslash mathit\textbraceleft{{GW}}\textbraceright\$ {{Calculations}} of {{Ground-State Properties}}},
  author = {{Garc{\'i}a-Gonz{\'a}lez}, P. and Godby, R. W.},
  year = 2002,
  month = jan,
  journal = {Phys. Rev. Lett.},
  volume = {88},
  number = {5},
  pages = {056406},
  publisher = {American Physical Society},
  doi = {10.1103/PhysRevLett.88.056406},
  urldate = {2026-01-27},
}

@article{godbySelfenergyOperatorsExchangecorrelation1988,
  title = {Self-Energy Operators and Exchange-Correlation Potentials in Semiconductors},
  author = {Godby, R. W. and Schl{\"u}ter, M. and Sham, L. J.},
  year = 1988,
  month = jun,
  journal = {Phys. Rev. B},
  volume = {37},
  number = {17},
  pages = {10159--10175},
  publisher = {American Physical Society},
  doi = {10.1103/PhysRevB.37.10159},
  urldate = {2026-01-27},
}

@article{golzeGWCompendiumPractical2019a,
  title = {The \emph{GW} Compendium: {{A Practical Guide}} to {{Theoretical Photoemission Spectroscopy}}},
  shorttitle = {The {{GW Compendium}}},
  author = {Golze, Dorothea and Dvorak, Marc and Rinke, Patrick},
  year = 2019,
  journal = {Front. Chem.},
  volume = {7},
  issn = {2296-2646},
  urldate = {2023-02-15},
}

@article{govoniLargeScaleGW2015,
  title = {Large {{Scale GW Calculations}}},
  author = {Govoni, Marco and Galli, Giulia},
  year = 2015,
  month = jun,
  journal = {J. Chem. Theory Comput.},
  volume = {11},
  number = {6},
  pages = {2680--2696},
  publisher = {American Chemical Society},
  issn = {1549-9618},
  doi = {10.1021/ct500958p},
  urldate = {2026-01-27},
}

@article{hanAnalyticContinuationPade2017,
  title = {Analytic {{Continuation}} with {{Pad\'e Decomposition}}},
  author = {Han, Xing-Jie and Liao, Hai-Jun and Xie, Hai-Dong and Huang, Rui-Zhen and Meng, Zi-Yang and Xiang, Tao},
  year = 2017,
  month = jul,
  journal = {Chinese Phys. Lett.},
  volume = {34},
  number = {7},
  pages = {077102},
  publisher = {{Chinese Physical Society and IOP Publishing Ltd}},
  issn = {0256-307X},
  doi = {10.1088/0256-307X/34/7/077102},
  urldate = {2023-03-23},
}

@article{harshaQuasiparticleFullySelfconsistent2024,
  title = {Quasiparticle and Fully Self-Consistent \emph{GW} Methods: {{An}} Unbiased Analysis Using {{Gaussian}} Orbitals},
  shorttitle = {Quasiparticle and Fully Self-Consistent \emph{GW} Methods},
  author = {Harsha, Gaurav and Abraham, Vibin and Wen, Ming and Zgid, Dominika},
  year = 2024,
  month = dec,
  journal = {Phys. Rev. B},
  volume = {110},
  number = {23},
  pages = {235146},
  publisher = {American Physical Society},
  doi = {10.1103/PhysRevB.110.235146},
  urldate = {2026-01-26},
}

@article{hedinNewMethodCalculating1965,
  title = {New {{Method}} for {{Calculating}} the {{One-Particle Green}}'s {{Function}} with {{Application}} to the {{Electron-Gas Problem}}},
  author = {Hedin, Lars},
  year = 1965,
  month = aug,
  journal = {Phys. Rev.},
  volume = {139},
  number = {3A},
  pages = {A796-A823},
  issn = {0031-899X},
  doi = {10.1103/PhysRev.139.A796},
  urldate = {2023-03-23},
}

@article{hirataTimedependentDensityFunctional1999,
  title = {Time-Dependent Density Functional Theory within the {{Tamm}}--{{Dancoff}} Approximation},
  author = {Hirata, So and {Head-Gordon}, Martin},
  year = 1999,
  month = dec,
  journal = {Chem. Phys. Lett.},
  volume = {314},
  number = {3},
  pages = {291--299},
  issn = {0009-2614},
  doi = {10.1016/S0009-2614(99)01149-5},
  urldate = {2026-03-03},
}

@article{hiroseAllelectron$GW$+BetheSalpeterCalculations2015,
  title = {All-Electron \emph{GW}+{{Bethe-Salpeter}} Calculations on Small Molecules},
  author = {Hirose, Daichi and Noguchi, Yoshifumi and Sugino, Osamu},
  year = 2015,
  month = may,
  journal = {Phys. Rev. B},
  volume = {91},
  number = {20},
  pages = {205111},
  publisher = {American Physical Society},
  doi = {10.1103/PhysRevB.91.205111},
  urldate = {2026-01-27},
}

@article{holmFullySelfconsistent$mathrmGW$1998b,
  title = {Fully Self-Consistent \emph{GW} Self-Energy of the Electron Gas},
  author = {Holm, B. and {von Barth}, U.},
  year = 1998,
  month = jan,
  journal = {Phys. Rev. B},
  volume = {57},
  number = {4},
  pages = {2108--2117},
  publisher = {American Physical Society},
  doi = {10.1103/PhysRevB.57.2108},
  urldate = {2026-01-27},
}

@article{huserQuasiparticleGWCalculations2013,
  title = {Quasiparticle {{GW}} Calculations for Solids, Molecules, and Two-Dimensional Materials},
  author = {H{\"u}ser, Falco and Olsen, Thomas and Thygesen, Kristian S.},
  year = 2013,
  month = jun,
  journal = {Phys. Rev. B},
  volume = {87},
  number = {23},
  pages = {235132},
  publisher = {American Physical Society},
  doi = {10.1103/PhysRevB.87.235132},
  urldate = {2026-01-27},
}

@article{hybertsenElectronCorrelationSemiconductors1986,
  title = {Electron Correlation in Semiconductors and Insulators: {{Band}} Gaps and Quasiparticle Energies},
  shorttitle = {Electron Correlation in Semiconductors and Insulators},
  author = {Hybertsen, Mark S. and Louie, Steven G.},
  year = 1986,
  month = oct,
  journal = {Phys. Rev. B},
  volume = {34},
  number = {8},
  pages = {5390--5413},
  publisher = {American Physical Society},
  doi = {10.1103/PhysRevB.34.5390},
  urldate = {2026-02-02},
}

@article{hybertsenFirstPrinciplesTheoryQuasiparticles1985,
  title = {First-{{Principles Theory}} of {{Quasiparticles}}: {{Calculation}} of {{Band Gaps}} in {{Semiconductors}} and {{Insulators}}},
  shorttitle = {First-{{Principles Theory}} of {{Quasiparticles}}},
  author = {Hybertsen, Mark S. and Louie, Steven G.},
  year = 1985,
  month = sep,
  journal = {Phys. Rev. Lett.},
  volume = {55},
  number = {13},
  pages = {1418--1421},
  publisher = {American Physical Society},
  doi = {10.1103/PhysRevLett.55.1418},
  urldate = {2026-01-27},
}

@article{iskakovGreenWeakCouplingImplementation2025,
  title = {Green/{{WeakCoupling}}: {{Implementation}} of Fully Self-Consistent Finite-Temperature Many-Body Perturbation Theory for Molecules and Solids},
  shorttitle = {Green/{{WeakCoupling}}},
  author = {Iskakov, Sergei and Yeh, Chia-Nan and Pokhilko, Pavel and Yu, Yang and Zhang, Lei and Harsha, Gaurav and Abraham, Vibin and Wen, Ming and Wang, Munkhorgil and Adamski, Jacob and Chen, Tianran and Gull, Emanuel and Zgid, Dominika},
  year = 2025,
  month = jan,
  journal = {Comput. Phys. Commun.},
  volume = {306},
  pages = {109380},
  issn = {0010-4655},
  doi = {10.1016/j.cpc.2024.109380},
  urldate = {2024-10-21},
}

@article{iskakovInitioSelfenergyEmbedding2020,
  title = {Ab Initio Self-Energy Embedding for the Photoemission Spectra of {{NiO}} and {{MnO}}},
  author = {Iskakov, Sergei and Yeh, Chia-Nan and Gull, Emanuel and Zgid, Dominika},
  year = 2020,
  month = aug,
  journal = {Phys. Rev. B},
  volume = {102},
  number = {8},
  pages = {085105},
  publisher = {American Physical Society},
  doi = {10.1103/PhysRevB.102.085105},
  urldate = {2026-03-25},
}

@misc{johnsonComputationalChemistryComparison2002a,
  title = {Computational {{Chemistry Comparison}} and {{Benchmark Database}}, {{NIST Standard Reference Database}} 101},
  author = {Johnson III, R. D.},
  year = 2002,
  publisher = {{National Institute of Standards and Technology}},
  doi = {10.18434/T47C7Z},
  urldate = {2026-01-27},
  copyright = {License Information for NIST data}
}

@article{kaplanQuasiParticleSelfConsistentGW2016a,
  title = {Quasi-{{Particle Self-Consistent \emph{GW}}} for {{Molecules}}},
  author = {Kaplan, F. and Harding, M. E. and Seiler, C. and Weigend, F. and Evers, F. and {van Setten}, M. J.},
  year = 2016,
  month = jun,
  journal = {J. Chem. Theory Comput.},
  volume = {12},
  number = {6},
  pages = {2528--2541},
  publisher = {American Chemical Society},
  issn = {1549-9618},
  doi = {10.1021/acs.jctc.5b01238},
  urldate = {2026-03-26},
}

@article{kendallElectronAffinitiesFirstrow1992,
  title = {Electron Affinities of the First-row Atoms Revisited. {{Systematic}} Basis Sets and Wave Functions},
  author = {Kendall, Rick A. and Dunning, Jr., Thom H. and Harrison, Robert J.},
  year = 1992,
  month = may,
  journal = {J. Chem. Phys.},
  volume = {96},
  number = {9},
  pages = {6796--6806},
  issn = {0021-9606},
  doi = {10.1063/1.462569},
  urldate = {2026-01-27},
}

@article{knyshReferenceCC3Excitation2024,
  title = {Reference {{CC3 Excitation Energies}} for {{Organic Chromophores}}: {{Benchmarking TD-DFT}}, {{BSE}}/{{GW}}, and {{Wave Function Methods}}},
  shorttitle = {Reference {{CC3 Excitation Energies}} for {{Organic Chromophores}}},
  author = {Knysh, Iryna and Lipparini, Filippo and Blondel, Aymeric and Duchemin, Ivan and Blase, Xavier and Loos, Pierre-Fran{\c c}ois and Jacquemin, Denis},
  year = 2024,
  month = sep,
  journal = {J. Chem. Theory Comput.},
  volume = {20},
  number = {18},
  pages = {8152--8174},
  publisher = {American Chemical Society},
  issn = {1549-9618},
  doi = {10.1021/acs.jctc.4c00906},
  urldate = {2026-01-26},
}

@article{kochCC3ModelIterative1997,
  title = {The {{CC3}} Model: {{An}} Iterative Coupled Cluster Approach Including Connected Triples},
  shorttitle = {The {{CC3}} Model},
  author = {Koch, Henrik and Christiansen, Ove and J{\o}rgensen, Poul and {Sanchez de Mer{\'a}s}, Alfredo M. and Helgaker, Trygve},
  year = 1997,
  month = feb,
  journal = {J. Chem. Phys.},
  volume = {106},
  number = {5},
  pages = {1808--1818},
  issn = {0021-9606},
  doi = {10.1063/1.473322},
  urldate = {2026-03-30},
}

@article{korzdorferStrategyFindingReliable2012,
  title = {Strategy for Finding a Reliable Starting Point for ${{G}}_ 0{{W}}_0$ Demonstrated for Molecules},
  author = {K{\"o}rzd{\"o}rfer, Thomas and Marom, Noa},
  year = 2012,
  month = jul,
  journal = {Phys. Rev. B},
  volume = {86},
  number = {4},
  pages = {041110},
  publisher = {American Physical Society},
  doi = {10.1103/PhysRevB.86.041110},
  urldate = {2025-08-26},
}

@article{krylovEquationofMotionCoupledClusterMethods2008b,
  title = {Equation-of-{{Motion Coupled-Cluster Methods}} for {{Open-Shell}} and {{Electronically Excited Species}}: {{The Hitchhiker}}'s {{Guide}} to {{Fock Space}}},
  shorttitle = {Equation-of-{{Motion Coupled-Cluster Methods}} for {{Open-Shell}} and {{Electronically Excited Species}}},
  author = {Krylov, Anna I.},
  year = 2008,
  month = may,
  journal = {Annu. Rev. Phys. Chem.},
  volume = {59},
  number = {Volume 59, 2008},
  pages = {433--462},
  publisher = {Annual Reviews},
  issn = {0066-426X, 1545-1593},
  doi = {10.1146/annurev.physchem.59.032607.093602},
  urldate = {2026-01-28},
}

@article{kutepovElectronicStructureNa2016b,
  title = {Electronic Structure of {{Na}}, {{K}}, {{Si}}, and {{LiF}} from Self-Consistent Solution of {{Hedin}}'s Equations Including Vertex Corrections},
  author = {Kutepov, Andrey L.},
  year = 2016,
  month = oct,
  journal = {Phys. Rev. B},
  volume = {94},
  number = {15},
  pages = {155101},
  publisher = {American Physical Society},
  doi = {10.1103/PhysRevB.94.155101},
  urldate = {2026-01-27},
}

@article{kutepovSelfconsistentSolutionHedins2017,
  title = {Self-Consistent Solution of {{Hedin}}'s Equations: {{Semiconductors}} and Insulators},
  shorttitle = {Self-Consistent Solution of {{Hedin}}'s Equations},
  author = {Kutepov, Andrey L.},
  year = 2017,
  month = may,
  journal = {Phys. Rev. B},
  volume = {95},
  number = {19},
  pages = {195120},
  publisher = {American Physical Society},
  doi = {10.1103/PhysRevB.95.195120},
  urldate = {2026-01-27},
}

@article{lanTestingSelfenergyEmbedding2017,
  title = {Testing Self-Energy Embedding Theory in Combination with {{GW}}},
  author = {Lan, Tran Nguyen and Shee, Avijit and Li, Jia and Gull, Emanuel and Zgid, Dominika},
  year = 2017,
  month = oct,
  journal = {Phys. Rev. B},
  volume = {96},
  number = {15},
  pages = {155106},
  publisher = {American Physical Society},
  doi = {10.1103/PhysRevB.96.155106},
  urldate = {2026-03-25},
}

@article{larsonRolePlasmonpoleModel2013,
  title = {Role of the Plasmon-Pole Model in the \emph{GW} Approximation},
  author = {Larson, Paul and Dvorak, Marc and Wu, Zhigang},
  year = 2013,
  month = sep,
  journal = {Phys. Rev. B},
  volume = {88},
  number = {12},
  pages = {125205},
  publisher = {American Physical Society},
  doi = {10.1103/PhysRevB.88.125205},
  urldate = {2026-03-26},
}

@article{liSparseSamplingApproach2020a,
  title = {Sparse Sampling Approach to Efficient Ab Initio Calculations at Finite Temperature},
  author = {Li, Jia and Wallerberger, Markus and Chikano, Naoya and Yeh, Chia-Nan and Gull, Emanuel and Shinaoka, Hiroshi},
  year = 2020,
  month = jan,
  journal = {Phys. Rev. B},
  volume = {101},
  number = {3},
  pages = {035144},
  publisher = {American Physical Society},
  doi = {10.1103/PhysRevB.101.035144},
  urldate = {2023-08-01},
}

@article{loosDynamicalCorrectionBethe2020,
  title = {Dynamical Correction to the {{Bethe}}--{{Salpeter}} Equation beyond the Plasmon-Pole Approximation},
  author = {Loos, Pierre-Fran{\c c}ois and Blase, Xavier},
  year = 2020,
  month = sep,
  journal = {J. Chem. Phys.},
  volume = {153},
  number = {11},
  pages = {114120},
  issn = {0021-9606, 1089-7690},
  doi = {10.1063/5.0023168},
  urldate = {2026-01-23},
}

@article{loosMountaineeringStrategyExcited2018,
  title = {A {{Mountaineering Strategy}} to {{Excited States}}: {{Highly Accurate Reference Energies}} and {{Benchmarks}}},
  shorttitle = {A {{Mountaineering Strategy}} to {{Excited States}}},
  author = {Loos, Pierre-Fran{\c c}ois and Scemama, Anthony and Blondel, Aymeric and Garniron, Yann and Caffarel, Michel and Jacquemin, Denis},
  year = 2018,
  month = aug,
  journal = {J. Chem. Theory Comput.},
  volume = {14},
  number = {8},
  pages = {4360--4379},
  publisher = {American Chemical Society},
  issn = {1549-9618},
  doi = {10.1021/acs.jctc.8b00406},
  urldate = {2026-01-27},
}

@article{loosStaticDynamicBethe2022,
  title = {Static and Dynamic {{Bethe}}--{{Salpeter}} Equations in the {{T-matrix}} Approximation},
  author = {Loos, Pierre-Fran{\c c}ois and Romaniello, Pina},
  year = 2022,
  month = apr,
  journal = {J. Chem. Phys.},
  volume = {156},
  number = {16},
  pages = {164101},
  issn = {0021-9606},
  doi = {10.1063/5.0088364},
  urldate = {2026-01-26},
}

@article{maExcitedStatesBiological2009,
  title = {Excited States of Biological Chromophores Studied Using Many-Body Perturbation Theory: {{Effects}} of Resonant-Antiresonant Coupling and Dynamical Screening},
  shorttitle = {Excited States of Biological Chromophores Studied Using Many-Body Perturbation Theory},
  author = {Ma, Yuchen and Rohlfing, Michael and Molteni, Carla},
  year = 2009,
  month = dec,
  journal = {Phys. Rev. B},
  volume = {80},
  number = {24},
  pages = {241405},
  publisher = {American Physical Society},
  doi = {10.1103/PhysRevB.80.241405},
  urldate = {2026-01-26},
}

@article{maggioGWVertexCorrected2017a,
  title = {{{\emph{GW} Vertex Corrected Calculations}} for {{Molecular Systems}}},
  author = {Maggio, Emanuele and Kresse, Georg},
  year = 2017,
  month = oct,
  journal = {J. Chem. Theory Comput.},
  volume = {13},
  number = {10},
  pages = {4765--4778},
  publisher = {American Chemical Society},
  issn = {1549-9618},
  doi = {10.1021/acs.jctc.7b00586},
  urldate = {2023-02-23},
}

@article{mejuto-zaeraAreMultiquasiparticleInteractions2021a,
  title = {Are Multi-Quasiparticle Interactions Important in Molecular Ionization?},
  author = {{Mejuto-Zaera}, Carlos and Weng, Guorong and Romanova, Mariya and Cotton, Stephen J. and Whaley, K. Birgitta and Tubman, Norm M. and Vl{\v c}ek, Vojt{\v e}ch},
  year = 2021,
  month = mar,
  journal = {J. Chem. Phys.},
  volume = {154},
  number = {12},
  pages = {121101},
  issn = {0021-9606, 1089-7690},
  doi = {10.1063/5.0044060},
  urldate = {2023-01-19},
}

@article{mesterChargeTransferExcitationsDensity2022,
  title = {Charge-{{Transfer Excitations}} within {{Density Functional Theory}}: {{How Accurate Are}} the {{Most Recommended Approaches}}?},
  shorttitle = {Charge-{{Transfer Excitations}} within {{Density Functional Theory}}},
  author = {Mester, D{\'a}vid and K{\'a}llay, Mih{\'a}ly},
  year = 2022,
  month = mar,
  journal = {J. Chem. Theory Comput.},
  volume = {18},
  number = {3},
  pages = {1646--1662},
  publisher = {American Chemical Society},
  issn = {1549-9618},
  doi = {10.1021/acs.jctc.1c01307},
  urldate = {2026-01-28},
}

@article{onidaElectronicExcitationsDensityfunctional2002a,
  title = {Electronic Excitations: Density-Functional versus Many-Body {{Green}}'s-Function Approaches},
  shorttitle = {Electronic Excitations},
  author = {Onida, Giovanni and Reining, Lucia and Rubio, Angel},
  year = 2002,
  month = jun,
  journal = {Rev. Mod. Phys.},
  volume = {74},
  number = {2},
  pages = {601--659},
  publisher = {American Physical Society},
  doi = {10.1103/RevModPhys.74.601},
  urldate = {2026-01-26},
}

@article{pritchardNewBasisSet2019,
  title = {New {{Basis Set Exchange}}: {{An Open}}, {{Up-to-Date Resource}} for the {{Molecular Sciences Community}}},
  shorttitle = {New {{Basis Set Exchange}}},
  author = {Pritchard, Benjamin P. and Altarawy, Doaa and Didier, Brett and Gibson, Tara D. and Windus, Theresa L.},
  year = 2019,
  month = nov,
  journal = {J. Chem. Inf. Model.},
  volume = {59},
  number = {11},
  pages = {4814--4820},
  issn = {1549-9596, 1549-960X},
  doi = {10.1021/acs.jcim.9b00725},
  urldate = {2026-01-27},
  copyright = {https://doi.org/10.15223/policy-029}
}

@article{purvisFullCoupledclusterSingles1982,
  title = {A Full Coupled-cluster Singles and Doubles Model: {{The}} Inclusion of Disconnected Triples},
  shorttitle = {A Full Coupled-cluster Singles and Doubles Model},
  author = {Purvis, III, George D. and Bartlett, Rodney J.},
  year = 1982,
  month = feb,
  journal = {J. Chem. Phys.},
  volume = {76},
  number = {4},
  pages = {1910--1918},
  issn = {0021-9606},
  doi = {10.1063/1.443164},
  urldate = {2026-01-27},
}

@article{rangelEvaluatingGWApproximation2016a,
  title = {Evaluating the {{GW Approximation}} with {{CCSD}}({{T}}) for {{Charged Excitations Across}} the {{Oligoacenes}}},
  author = {Rangel, Tonatiuh and Hamed, Samia M. and Bruneval, Fabien and Neaton, Jeffrey B.},
  year = 2016,
  month = jun,
  journal = {J. Chem. Theory Comput.},
  volume = {12},
  number = {6},
  pages = {2834--2842},
  publisher = {American Chemical Society},
  issn = {1549-9618},
  doi = {10.1021/acs.jctc.6b00163},
  urldate = {2026-03-26},
}

@article{reiningGWApproximationContent2018,
  title = {The \emph{GW} Approximation: Content, Successes and Limitations},
  shorttitle = {The {{GW}} Approximation},
  author = {Reining, Lucia},
  year = 2018,
  journal = {WIREs Comput. Mol. Sci.},
  volume = {8},
  number = {3},
  pages = {e1344},
  issn = {1759-0884},
  doi = {10.1002/wcms.1344},
  urldate = {2026-02-02},
  copyright = {\copyright{} 2017 Wiley Periodicals, Inc.},
}

@article{ren$GW$ApproximationSecondorder2015,
  title = {Beyond the \emph{GW} Approximation: {{A}} Second-Order Screened Exchange Correction},
  shorttitle = {Beyond the \emph{GW} Approximation},
  author = {Ren, Xinguo and Marom, Noa and Caruso, Fabio and Scheffler, Matthias and Rinke, Patrick},
  year = 2015,
  month = aug,
  journal = {Phys. Rev. B},
  volume = {92},
  number = {8},
  pages = {081104},
  publisher = {American Physical Society},
  doi = {10.1103/PhysRevB.92.081104},
  urldate = {2026-03-26},
}

@article{renResolutionofidentityApproachHartree2012,
  title = {Resolution-of-Identity Approach to {{Hartree}}--{{Fock}}, Hybrid Density Functionals, {{RPA}}, {{MP2}} and {{GW}} with Numeric Atom-Centered Orbital Basis Functions},
  author = {Ren, Xinguo and Rinke, Patrick and Blum, Volker and Wieferink, J{\"u}rgen and Tkatchenko, Alexandre and Sanfilippo, Andrea and Reuter, Karsten and Scheffler, Matthias},
  year = 2012,
  month = may,
  journal = {New J. Phys.},
  volume = {14},
  number = {5},
  pages = {053020},
  publisher = {IOP Publishing},
  issn = {1367-2630},
  doi = {10.1088/1367-2630/14/5/053020},
  urldate = {2026-01-26},
}

@article{rohlfingElectronholeExcitationsOptical2000,
  title = {Electron-Hole Excitations and Optical Spectra from First Principles},
  author = {Rohlfing, Michael and Louie, Steven G.},
  year = 2000,
  month = aug,
  journal = {Phys. Rev. B},
  volume = {62},
  number = {8},
  pages = {4927--4944},
  issn = {0163-1829, 1095-3795},
  doi = {10.1103/PhysRevB.62.4927},
  urldate = {2026-02-02},
  copyright = {http://link.aps.org/licenses/aps-default-license},
}

@article{rohlfingElectronHoleExcitationsSemiconductors1998,
  title = {Electron-{{Hole Excitations}} in {{Semiconductors}} and {{Insulators}}},
  author = {Rohlfing, Michael and Louie, Steven G.},
  year = 1998,
  month = sep,
  journal = {Phys. Rev. Lett.},
  volume = {81},
  number = {11},
  pages = {2312--2315},
  issn = {0031-9007, 1079-7114},
  doi = {10.1103/PhysRevLett.81.2312},
  urldate = {2026-02-02},
  copyright = {http://link.aps.org/licenses/aps-default-license},
}

@article{romanielloDoubleExcitationsFinite2009,
  title = {Double Excitations in Finite Systems},
  author = {Romaniello, P. and Sangalli, D. and Berger, J. A. and Sottile, F. and Molinari, L. G. and Reining, L. and Onida, G.},
  year = 2009,
  month = jan,
  journal = {J. Chem. Phys.},
  volume = {130},
  number = {4},
  pages = {044108},
  issn = {0021-9606},
  doi = {10.1063/1.3065669},
  urldate = {2026-01-26},
}

@article{rostgaardFullySelfconsistentGW2010b,
  title = {Fully Self-Consistent {{GW}} Calculations for Molecules},
  author = {Rostgaard, C. and Jacobsen, K. W. and Thygesen, K. S.},
  year = 2010,
  month = feb,
  journal = {Phys. Rev. B},
  volume = {81},
  number = {8},
  pages = {085103},
  publisher = {American Physical Society},
  doi = {10.1103/PhysRevB.81.085103},
  urldate = {2026-02-02},
}

@article{rubioExcitedStatesWater2008,
  title = {Excited States of the Water Molecule: {{Analysis}} of the Valence and {{Rydberg}} Character},
  shorttitle = {Excited States of the Water Molecule},
  author = {Rubio, Mercedes and {Serrano-Andr{\'e}s}, Luis and Merch{\'a}n, Manuela},
  year = 2008,
  month = mar,
  journal = {J. Chem. Phys.},
  volume = {128},
  number = {10},
  pages = {104305},
  issn = {0021-9606},
  doi = {10.1063/1.2837827},
  urldate = {2026-03-30},
}

@article{salpeterRelativisticEquationBoundState1951,
  title = {A {{Relativistic Equation}} for {{Bound-State Problems}}},
  author = {Salpeter, E. E. and Bethe, H. A.},
  year = 1951,
  month = dec,
  journal = {Phys. Rev.},
  volume = {84},
  number = {6},
  pages = {1232--1242},
  publisher = {American Physical Society},
  doi = {10.1103/PhysRev.84.1232},
  urldate = {2026-01-26},
}

@article{sangalliDoubleExcitationsCorrelated2011,
  title = {Double Excitations in Correlated Systems: {{A}} Many--Body Approach},
  shorttitle = {Double Excitations in Correlated Systems},
  author = {Sangalli, Davide and Romaniello, Pina and Onida, Giovanni and Marini, Andrea},
  year = 2011,
  month = jan,
  journal = {J. Chem. Phys.},
  volume = {134},
  number = {3},
  pages = {034115},
  issn = {0021-9606},
  doi = {10.1063/1.3518705},
  urldate = {2026-01-27},
}

@article{shinaokaCompressingGreensFunction2017a,
  title = {Compressing {{Green}}'s Function Using Intermediate Representation between Imaginary-Time and Real-Frequency Domains},
  author = {Shinaoka, Hiroshi and Otsuki, Junya and Ohzeki, Masayuki and Yoshimi, Kazuyoshi},
  year = 2017,
  month = jul,
  journal = {Phys. Rev. B},
  volume = {96},
  number = {3},
  pages = {035147},
  publisher = {American Physical Society},
  doi = {10.1103/PhysRevB.96.035147},
  urldate = {2026-01-27},
}

@article{shishkinSelfconsistent$GW$Calculations2007a,
  title = {Self-Consistent \emph{GW} Calculations for Semiconductors and Insulators},
  author = {Shishkin, M. and Kresse, G.},
  year = 2007,
  month = jun,
  journal = {Phys. Rev. B},
  volume = {75},
  number = {23},
  pages = {235102},
  publisher = {American Physical Society},
  doi = {10.1103/PhysRevB.75.235102},
  urldate = {2026-01-27},
}

@article{stanFullySelfconsistentGW2006,
  title = {Fully Self-Consistent {{GW}} Calculations for Atoms and Molecules},
  author = {Stan, A. and Dahlen, N. E. and van Leeuwen, R.},
  year = 2006,
  month = sep,
  journal = {EPL},
  volume = {76},
  number = {2},
  pages = {298},
  publisher = {IOP Publishing},
  issn = {0295-5075},
  doi = {10.1209/epl/i2006-10266-6},
  urldate = {2026-02-02},
}

@article{strinatiApplicationGreensFunctions1988,
  title = {Application of the {{Green}}'s Functions Method to the Study of the Optical Properties of Semiconductors},
  author = {Strinati, G.},
  year = 1988,
  month = dec,
  journal = {Riv. Nuovo Cim.},
  volume = {11},
  number = {12},
  pages = {1--86},
  issn = {1826-9850},
  doi = {10.1007/BF02725962},
  urldate = {2026-01-26},
}

@article{strinatiEffectsDynamicalScreening1984,
  title = {Effects of Dynamical Screening on Resonances at Inner-Shell Thresholds in Semiconductors},
  author = {Strinati, G.},
  year = 1984,
  month = may,
  journal = {Phys. Rev. B},
  volume = {29},
  number = {10},
  pages = {5718--5726},
  publisher = {American Physical Society},
  doi = {10.1103/PhysRevB.29.5718},
  urldate = {2026-02-02},
}

@article{sunLibcintEfficientGeneral2015b,
  title = {Libcint: {{An}} Efficient General Integral Library for {{Gaussian}} Basis Functions},
  shorttitle = {Libcint},
  author = {Sun, Qiming},
  year = 2015,
  journal = {J. Comput. Chem.},
  volume = {36},
  number = {22},
  pages = {1664--1671},
  issn = {1096-987X},
  doi = {10.1002/jcc.23981},
  urldate = {2026-01-27},
  copyright = {\copyright{} 2015 Wiley Periodicals, Inc.},
}

@article{sunPySCFPythonbasedSimulations2018a,
  title = {{{PySCF}}: The {{Python-based}} Simulations of Chemistry Framework},
  shorttitle = {{{PySCF}}},
  author = {Sun, Qiming and Berkelbach, Timothy C. and Blunt, Nick S. and Booth, George H. and Guo, Sheng and Li, Zhendong and Liu, Junzi and McClain, James D. and Sayfutyarova, Elvira R. and Sharma, Sandeep and Wouters, Sebastian and Chan, Garnet Kin-Lic},
  year = 2018,
  journal = {WIREs Comput. Mol. Sci.},
  volume = {8},
  number = {1},
  pages = {e1340},
  issn = {1759-0884},
  doi = {10.1002/wcms.1340},
  urldate = {2023-11-09},
  copyright = {\copyright{} 2017 Wiley Periodicals, Inc.},
}

@article{sunRecentDevelopmentsPySCF2020a,
  title = {Recent Developments in the {{PySCF}} Program Package},
  author = {Sun, Qiming and Zhang, Xing and Banerjee, Samragni and Bao, Peng and Barbry, Marc and Blunt, Nick S. and Bogdanov, Nikolay A. and Booth, George H. and Chen, Jia and Cui, Zhi-Hao and Eriksen, Janus J. and Gao, Yang and Guo, Sheng and Hermann, Jan and Hermes, Matthew R. and Koh, Kevin and Koval, Peter and Lehtola, Susi and Li, Zhendong and Liu, Junzi and Mardirossian, Narbe and McClain, James D. and Motta, Mario and Mussard, Bastien and Pham, Hung Q. and Pulkin, Artem and Purwanto, Wirawan and Robinson, Paul J. and Ronca, Enrico and Sayfutyarova, Elvira R. and Scheurer, Maximilian and Schurkus, Henry F. and Smith, James E. T. and Sun, Chong and Sun, Shi-Ning and Upadhyay, Shiv and Wagner, Lucas K. and Wang, Xiao and White, Alec and Whitfield, James Daniel and Williamson, Mark J. and Wouters, Sebastian and Yang, Jun and Yu, Jason M. and Zhu, Tianyu and Berkelbach, Timothy C. and Sharma, Sandeep and Sokolov, Alexander Yu. and Chan, Garnet Kin-Lic},
  year = 2020,
  month = jul,
  journal = {J. Chem. Phys.},
  volume = {153},
  number = {2},
  pages = {024109},
  issn = {0021-9606},
  doi = {10.1063/5.0006074},
  urldate = {2023-11-09},
}

@book{szaboModernQuantumChemistry1996,
  title = {Modern {{Quantum Chemistry}}: {{Introduction}} to {{Advanced Electronic Structure Theory}}},
  shorttitle = {Modern {{Quantum Chemistry}}},
  author = {Szabo, Attila and Ostlund, Neil S.},
  year = 1996,
  month = jul,
  publisher = {Courier Corporation},
  googlebooks = {6mV9gYzEkgIC},
  isbn = {978-0-486-69186-2},
}

@article{vanschilfgaardeQuasiparticleSelfConsistent$GW$2006,
  title = {Quasiparticle {{Self-Consistent}} \emph{GW} {{Theory}}},
  author = {{van~Schilfgaarde}, M. and Kotani, Takao and Faleev, S.},
  year = 2006,
  month = jun,
  journal = {Phys. Rev. Lett.},
  volume = {96},
  number = {22},
  pages = {226402},
  publisher = {American Physical Society},
  doi = {10.1103/PhysRevLett.96.226402},
  urldate = {2024-04-05},
}

@article{vansettenGW100BenchmarkingG0W02015b,
  title = {{{GW100}}: {{Benchmarking G0W0}} for {{Molecular Systems}}},
  shorttitle = {{{GW100}}},
  author = {{van Setten}, Michiel J. and Caruso, Fabio and Sharifzadeh, Sahar and Ren, Xinguo and Scheffler, Matthias and Liu, Fang and Lischner, Johannes and Lin, Lin and Deslippe, Jack R. and Louie, Steven G. and Yang, Chao and Weigend, Florian and Neaton, Jeffrey B. and Evers, Ferdinand and Rinke, Patrick},
  year = 2015,
  month = dec,
  journal = {J. Chem. Theory Comput.},
  volume = {11},
  number = {12},
  pages = {5665--5687},
  publisher = {American Chemical Society},
  issn = {1549-9618},
  doi = {10.1021/acs.jctc.5b00453},
  urldate = {2023-10-19},
}

@article{vidbergSolvingEliashbergEquations1977,
  title = {Solving the {{Eliashberg}} Equations by Means {{ofN-point Pad\'e}} Approximants},
  author = {Vidberg, H. J. and Serene, J. W.},
  year = 1977,
  month = nov,
  journal = {J. Low Temp. Phys.},
  volume = {29},
  number = {3},
  pages = {179--192},
  issn = {1573-7357},
  doi = {10.1007/BF00655090},
  urldate = {2026-01-26},
}

@article{vlcekStochasticVertexCorrections2019a,
  title = {Stochastic {{Vertex Corrections}}: {{Linear Scaling Methods}} for {{Accurate Quasiparticle Energies}}},
  shorttitle = {Stochastic {{Vertex Corrections}}},
  author = {Vl{\v c}ek, Vojt{\v e}ch},
  year = 2019,
  month = nov,
  journal = {J. Chem. Theory Comput.},
  volume = {15},
  number = {11},
  pages = {6254--6266},
  issn = {1549-9618, 1549-9626},
  doi = {10.1021/acs.jctc.9b00317},
  urldate = {2023-02-27},
}

@article{wangExcitonsSolidsPeriodic2020,
  title = {Excitons in {{Solids}} from {{Periodic Equation-of-Motion Coupled-Cluster Theory}}},
  author = {Wang, Xiao and Berkelbach, Timothy C.},
  year = 2020,
  month = may,
  journal = {J. Chem. Theory Comput.},
  volume = {16},
  number = {5},
  pages = {3095--3103},
  publisher = {American Chemical Society},
  issn = {1549-9618},
  doi = {10.1021/acs.jctc.0c00101},
  urldate = {2026-01-28},
}

@article{wenComparingSelfConsistentGW2024,
  title = {Comparing {{Self-Consistent ${GW}$}} and Vertex-Corrected ${{G_0W_0}}$ (${{G_0W_0\Gamma}}$) {{Accuracy}} for {{Molecular Ionization Potentials}}},
  author = {Wen, Ming and Abraham, Vibin and Harsha, Gaurav and Shee, Avijit and Whaley, K. Birgitta and Zgid, Dominika},
  year = 2024,
  month = apr,
  journal = {J. Chem. Theory Comput.},
  volume = {20},
  number = {8},
  pages = {3109--3120},
  publisher = {American Chemical Society},
  issn = {1549-9618},
  doi = {10.1021/acs.jctc.3c01279},
  urldate = {2026-01-26},
}

@misc{wenGreenbsePaperreferencebsescgw2026,
  title = {green-bse/paper-reference-bse-scgw},
  author = {Wen, Ming and Harsha, Gaurav and Zgid, Dominika},
  year = 2026,
  month = apr,
  doi = {10.5281/zenodo.19716094},
  urldate = {2026-04-24},
  howpublished = {Zenodo}
}

@article{wernerFastLinearScaling2003,
  title = {Fast Linear Scaling Second-Order {{M\o ller-Plesset}} Perturbation Theory ({{MP2}}) Using Local and Density Fitting Approximations},
  author = {Werner, Hans-Joachim and Manby, Frederick R. and Knowles, Peter J.},
  year = 2003,
  month = may,
  journal = {J. Chem. Phys.},
  volume = {118},
  number = {18},
  pages = {8149--8160},
  issn = {0021-9606},
  doi = {10.1063/1.1564816},
  urldate = {2026-01-27},
}

@article{yaoAllElectronBSEGW2022,
  title = {All-{{Electron BSE}}@{{{\emph{GW}}}} {{Method}} for {{{\emph{K}}}} -{{Edge Core Electron Excitation Energies}}},
  author = {Yao, Yi and Golze, Dorothea and Rinke, Patrick and Blum, Volker and Kanai, Yosuke},
  year = 2022,
  month = mar,
  journal = {J. Chem. Theory Comput.},
  volume = {18},
  number = {3},
  pages = {1569--1583},
  issn = {1549-9618, 1549-9626},
  doi = {10.1021/acs.jctc.1c01180},
  urldate = {2026-01-23},
  copyright = {https://doi.org/10.15223/policy-029},
}

@article{yeFastPeriodicGaussian2021,
  title = {Fast Periodic {{Gaussian}} Density Fitting by Range Separation},
  author = {Ye, Hong-Zhou and Berkelbach, Timothy C.},
  year = 2021,
  month = apr,
  journal = {J. Chem. Phys.},
  volume = {154},
  number = {13},
  pages = {131104},
  issn = {0021-9606},
  doi = {10.1063/5.0046617},
  urldate = {2026-01-27},
}

@article{yehFullySelfconsistentFinitetemperature2022a,
  title = {Fully Self-Consistent Finite-Temperature {{{\emph{GW}}}} in {{Gaussian Bloch}} Orbitals for Solids},
  author = {Yeh, Chia-Nan and Iskakov, Sergei and Zgid, Dominika and Gull, Emanuel},
  year = 2022,
  month = dec,
  journal = {Phys. Rev. B},
  volume = {106},
  number = {23},
  pages = {235104},
  publisher = {American Physical Society},
  doi = {10.1103/PhysRevB.106.235104},
  urldate = {2023-08-01},
}

@article{yehRelativisticSelfconsistent$GW$2022,
  title = {Relativistic Self-Consistent \emph{GW}: {{Exact}} Two-Component Formalism with One-Electron Approximation for Solids},
  shorttitle = {Relativistic Self-Consistent \emph{GW}},
  author = {Yeh, Chia-Nan and Shee, Avijit and Sun, Qiming and Gull, Emanuel and Zgid, Dominika},
  year = 2022,
  month = aug,
  journal = {Phys. Rev. B},
  volume = {106},
  number = {8},
  pages = {085121},
  publisher = {American Physical Society},
  doi = {10.1103/PhysRevB.106.085121},
  urldate = {2026-03-25},
}

@article{zhangDynamicalSecondorderBetheSalpeter2013,
  title = {Dynamical Second-Order {{Bethe-Salpeter}} Equation Kernel: {{A}} Method for Electronic Excitation beyond the Adiabatic Approximation},
  shorttitle = {Dynamical Second-Order {{Bethe-Salpeter}} Equation Kernel},
  author = {Zhang, Du and Steinmann, Stephan N. and Yang, Weitao},
  year = 2013,
  month = oct,
  journal = {J. Chem. Phys.},
  volume = {139},
  number = {15},
  pages = {154109},
  issn = {0021-9606},
  doi = {10.1063/1.4824907},
  urldate = {2026-02-02},
}

@article{zhangMinimalPoleRepresentation2024,
  title = {Minimal Pole Representation and Analytic Continuation of Matrix-Valued Correlation Functions},
  author = {Zhang, Lei and Yu, Yang and Gull, Emanuel},
  year = 2024,
  month = dec,
  journal = {Phys. Rev. B},
  volume = {110},
  number = {23},
  pages = {235131},
  publisher = {American Physical Society},
  doi = {10.1103/PhysRevB.110.235131},
  urldate = {2026-03-26},
}

@article{blaseFirstprinciples$mathitGW$Calculations2011,
  title = {First-Principles ${{GW}}$ Calculations for Fullerenes, Porphyrins, Phtalocyanine, and Other Molecules of Interest for Organic Photovoltaic Applications},
  author = {Blase, X. and Attaccalite, C. and Olevano, V.},
  year = 2011,
  month = mar,
  journal = {Phys. Rev. B},
  volume = {83},
  number = {11},
  pages = {115103},
  publisher = {American Physical Society},
  doi = {10.1103/PhysRevB.83.115103},
  urldate = {2026-04-13}
}

@article{brunevalBenchmarkingStartingPoints2013b,
  title = {Benchmarking the {{Starting Points}} of the {{GW Approximation}} for {{Molecules}}},
  author = {Bruneval, Fabien and Marques, Miguel A. L.},
  year = 2013,
  month = jan,
  journal = {J. Chem. Theory Comput.},
  volume = {9},
  number = {1},
  pages = {324--329},
  publisher = {American Chemical Society},
  issn = {1549-9618},
  doi = {10.1021/ct300835h},
  urldate = {2026-04-13}
}

@article{brunevalIonizationEnergyAtoms2012,
  title = {Ionization Energy of Atoms Obtained from {{GW}} Self-Energy or from Random Phase Approximation Total Energies},
  author = {Bruneval, Fabien},
  year = 2012,
  month = may,
  journal = {J. Chem. Phys.},
  volume = {136},
  number = {19},
  pages = {194107},
  issn = {0021-9606},
  doi = {10.1063/1.4718428},
  urldate = {2026-04-13}
}

@article{knightAccurateIonizationPotentials2016b,
  title = {Accurate {{Ionization Potentials}} and {{Electron Affinities}} of {{Acceptor Molecules III}}: {{A Benchmark}} of {{GW Methods}}},
  shorttitle = {Accurate {{Ionization Potentials}} and {{Electron Affinities}} of {{Acceptor Molecules III}}},
  author = {Knight, Joseph W. and Wang, Xiaopeng and Gallandi, Lukas and Dolgounitcheva, Olga and Ren, Xinguo and Ortiz, J. Vincent and Rinke, Patrick and K{\"o}rzd{\"o}rfer, Thomas and Marom, Noa},
  year = 2016,
  month = feb,
  journal = {J. Chem. Theory Comput.},
  volume = {12},
  number = {2},
  pages = {615--626},
  publisher = {American Chemical Society},
  issn = {1549-9618},
  doi = {10.1021/acs.jctc.5b00871},
  urldate = {2026-04-13}
}

@article{lewisVertexCorrectionsPolarizability2019b,
  title = {Vertex {{Corrections}} to the {{Polarizability Do Not Improve}} the {{GW Approximation}} for the {{Ionization Potential}} of {{Molecules}}},
  author = {Lewis, Alan M. and Berkelbach, Timothy C.},
  year = 2019,
  month = may,
  journal = {J. Chem. Theory Comput.},
  volume = {15},
  number = {5},
  pages = {2925--2932},
  publisher = {American Chemical Society},
  issn = {1549-9618},
  doi = {10.1021/acs.jctc.8b00995},
  urldate = {2026-04-13}
}

@article{vansettenGWMethodQuantumChemistry2013,
  title = {The {{GW-Method}} for {{Quantum Chemistry Applications}}: {{Theory}} and {{Implementation}}},
  shorttitle = {The {{GW-Method}} for {{Quantum Chemistry Applications}}},
  author = {{van Setten}, M. J. and Weigend, F. and Evers, F.},
  year = 2013,
  month = jan,
  journal = {J. Chem. Theory Comput.},
  volume = {9},
  number = {1},
  pages = {232--246},
  publisher = {American Chemical Society},
  issn = {1549-9618},
  doi = {10.1021/ct300648t},
  urldate = {2026-04-13}
}

@article{wangAssessingG0W0G01Approach2021,
  title = {Assessing the {{G0W0$\Gamma$0}}(1) {{Approach}}: {{Beyond G0W0}} with {{Hedin}}'s {{Full Second-Order Self-Energy Contribution}}},
  shorttitle = {Assessing the {{G0W0$\Gamma$0}}(1) {{Approach}}},
  author = {Wang, Yanyong and Rinke, Patrick and Ren, Xinguo},
  year = 2021,
  month = aug,
  journal = {J. Chem. Theory Comput.},
  volume = {17},
  number = {8},
  pages = {5140--5154},
  publisher = {American Chemical Society},
  issn = {1549-9618},
  doi = {10.1021/acs.jctc.1c00488},
  urldate = {2026-04-13}
}

@article{brunevalSystematicBenchmarkInitio2015,
  title = {A Systematic Benchmark of the Ab Initio {{Bethe-Salpeter}} Equation Approach for Low-Lying Optical Excitations of Small Organic Molecules},
  author = {Bruneval, Fabien and Hamed, Samia M. and Neaton, Jeffrey B.},
  year = 2015,
  month = jun,
  journal = {J. Chem. Phys.},
  volume = {142},
  number = {24},
  pages = {244101},
  issn = {0021-9606},
  doi = {10.1063/1.4922489},
  urldate = {2026-04-24}
}

@article{PhysRevB.100.085112,
  title = {Effect of propagator renormalization on the band gap of insulating solids},
  author = {Iskakov, Sergei and Rusakov, Alexander A. and Zgid, Dominika and Gull, Emanuel},
  journal = {Phys. Rev. B},
  volume = {100},
  issue = {8},
  pages = {085112},
  numpages = {7},
  year = {2019},
  month = {Aug},
  publisher = {American Physical Society},
  doi = {10.1103/PhysRevB.100.085112},
  url = {https://link.aps.org/doi/10.1103/PhysRevB.100.085112}
}

\end{document}